# Interplay of spin-lattice and electronic coupling far above Neel ordering in 2D antiferromagnetic $CrPS_4$ and its interface manifestation

Divya Jangra[1], Binoy Krishna De[2]*, Mukesh Kumar Dasoundhi[3], Sourav Chowdhury[4], Pragati Sharma[1], Kartick Biswas[2], Arup Basak[2], Dheeraj Kumar Gupta[1], Koushik chakraborty[1], Praveen Kumar Velpula[1], Pavan Nukala[2], U. Chandni[5], Arvind Kumar yogi[1], Mukul Gupta[1], Markus Hücker[3], Vasant G. Sathe[1]*

[1]UGC-DAE Consortium for Scientific Research, D.A. University Campus, Khandwa Road, Indore-452001, India

[2]Centre for Nano Science and Engineering, Indian Institute of Science, Bangalore-560012, India

[3]Department of Condensed Matter Physics, Weizmann Institute of Science, Rehovot, Israel

[4]Deutsches Elektronen-Synchrotron DESY, Notkestrasse 85, 22607 Hamburg, Germany

[5]Department of Instrumentation and Applied Physics, Indian Institute of Science, Bangalore-560012, India

## Abstract

A short-range spin correlation-driven, strongly intercoupled spin-phonon-electronic state far above $T_N$ (~38K) is identified in the low-dimensional van der Waals antiferromagnet $CrPS_4$. Temperature-dependent Raman spectroscopy reveals spin-phonon coupling persisting up to T* (~120 K), concomitant with local lattice distortion. The setting of vibrionic progression in photoluminescence spectra suggests the strengthening of electron-phonon coupling around T*. Furthermore, both the electrical transport and optoelectronic response also change significantly at T*. The results indicate that spin–phonon coupling above $T_N$ in $CrPS_4$ originates from local lattice distortions induced by short-range magnetic correlations, which in turn enhance the electron–phonon interaction. Furthermore, using a $CrPS_4/In_2Se_3$ heterostructure, we demonstrate that the anomaly associated with the coupled degrees of freedom in $CrPS_4$ also influences the adjacent $In_2Se_3$ layer. The lattice dynamics of $In_2Se_3$ is significantly modified across the magnetic anomaly of $CrPS_4$, and the coupled dynamics is observed at T*. This interfacial manifestation opens up new possibilities for achieving correlated multifunctionalities in artificially designed heterostructure.

## 1. Introduction

The observation of magnetism in two-dimensional (2D) materials has opened a new avenue for tuning functional properties. Unlike conventional materials, magnetism in these systems can be modulated not only by external magnetic fields but also by electric fields, optical fields, or strain [[1],[2],[3]]. Furthermore, the capacity to modify multifunctional properties via thickness variation, combined with inherent structural anisotropy, makes these materials exceptionally promising for future technologies [[4]]. However, the coupling between magnetic structure and the electronic states is remarkably complex, presenting significant challenges in tuning and stabilising electronic properties. Even minor variations in structural strain can profoundly alter both the electronic structure and the magnetic behaviour [[5]]. Consequently, research into the coupling mechanisms between magnetism and electronic structure in 2D materials has accelerated recently, with spin-phonon and electron-phonon interactions emerging as pivotal factors [[6]].

The correlation between multiple degrees of freedom provides a fertile platform for realizing multifunctional properties and next-generation device concepts [[7],[8]]. Interactions among spin, charge, lattice, and polarization degrees of freedom give rise to several exotic phenomena such as magnetoelectricity, multiferroicity, spin Hall effect, and the spin Seebeck effect [8,[9],[10]]. In this context, spin–charge–lattice coupling in two-dimensional (2D) materials is significantly enhanced owing to their structural and electronic anisotropy, reduced dimensionality, device compatibility and ability to form heterostructures with functionalities inaccessible in bulk counterparts [11,12]. Beyond conventional exchange interactions, perturbations of the local crystal field and spin–orbit coupling provide additional pathways for controlling spin–lattice interactions and spin-based electronic states [13,14,15].

Two-dimensional magnetic semiconductor materials, including $CrI_3$ [16], $CrBr_3$ [17], $CrCl_3$ [18], and Fe-based van der Waals magnets [19,20,21], are studied to investigate dimensionality-dependent magnetism and spin-lattice coupling [22]. However, $CrPS_4$ is of particular interest, owing to its comparatively high environmental stability [23]. In this regard, the magnetic structure and in-plane anisotropy make $CrPS_4$ a fascinating platform for exploring 2D magnetism and associated spin-lattice coupling. $CrPS_4$ is an A-type antiferromagnetic semiconductor having a Neel temperature ($T_N \sim 38$ K) [24]. The Magnetic order in this material is driven by magnetic exchange interactions and is closely intertwined with its lattice, electronic, and transport properties, highlighting the relevance of spin–phonon–electron interactions [25,26].

Although the magnetic structure and associated spin-phonon coupling near $T_N$ are well-studied [27,28], temperature-dependent structural and optical measurements revealed pronounced anomalies in $CrPS_4$ far above $T_N$, around 120–140 K [29,30] which remain unaddressed. In particular, an anomalous evolution of the *b*-lattice parameter in this temperature range suggests a local structural rearrangement [29]. Budnaik et al [30] investigated the optical response of this compound, demonstrating that the fundamental absorption edge corresponding to the bandgap evolution from 1.30 eV at 300 K to 1.48 eV at 15 K exhibits a change in the slope of its temperature dependence around 120-140 K. Furthermore, the dual-band photoluminescence of $CrPS_4$, previously associated with competing fluorescence and phosphorescence pathways, indicating fluorescence behaviour dominates above 130 K [31]. Consequently, the microscopic origin of this higher-temperature anomaly (around 120-140 K, we designated as T*) and its possible connection to the coupled spin, lattice, and electronic degrees of freedom remain unexplored.

Additionally, $CrPS_4$ has attracted interest because of strong proximity effects when brought in contact with other materials [32], specifically its unique magnon transport characteristics [33]. Constructing artificial van der Waals heterostructures from multiple functional materials provides a promising approach for examining such coupled degrees of freedom extending across an interface [34,35]. Here, we fabricate a heterostructure based on $CrPS_4$/α-$In_2Se_3$ and investigate how the spin-lattice coupling in antiferromagnetic $CrPS_4$ influences the adjacent nonmagnetic α-$In_2Se_3$. In particular, α-$In_2Se_3$ is a two-dimensional ferroelectric material, which can modify the interfacial electrostatic and band alignment [36]. Such interfacial coupling provides a route toward combining magnetic and ferroelectric functionalities within a single heterostructure [37].

Here, we systematically investigate the interplay among spin, charge, and lattice degrees of freedom in 2D $CrPS_4$ using temperature-dependent Raman measurements and optical spectroscopy, alongside magnetic and electrical characterizations. Systematic analysis of the phonon modes via Raman spectroscopy reveals deviations from their expected anharmonic behavior near the magnetic ordering temperature $T_N$, as well as at T*. Similarly, photoluminescence, electrical transport, and optoelectronic responses exhibit dramatic changes near T*. Analysis of the Raman modes suggested the presence of short-range ordering and a concomitant octahedral distortion of $CrS_6$ octahedra, which subsequently modifies the optical properties. Furthermore, we leverage the spin-phonon coupling in $CrPS_4$ at T* to modulate the phonon structure of the adjacent ferroelectric α-$In_2Se_3$ layer. These results establish a robust framework for understanding coupled degrees of freedom in $CrPS_4$ and highlight the potential of such interactions for engineering van der Waals heterostructures with tailored multifunctional properties [34].

## 2. Experimental section:

**Crystal Growth:** $CrPS_4$ single crystals were grown by the chemical vapor transport (CVT) method. Chromium powder (99.99%), red phosphorus powder (99.999%), and sulphur powder (99.999%) were weighed in a 1:1:4 molar ratio, with an additional 4% excess sulphur, and thoroughly mixed in an Ar glovebox. The powder mixture was then loaded into a quartz ampoule, which was evacuated to approximately $10^{-5}$ Torr and sealed. The ampoule was subsequently placed in a two-zone furnace, with the source and growth zones maintained at 650 °C and 550 °C, respectively, for 10 days. The furnace was then allowed to cool naturally to room temperature. Black, shiny, plate-like $CrPS_4$ single crystals were obtained at the growth end (550 °C) of the ampoule.

**X-Ray diffraction:** The X-ray diffraction measurements were carried out on a bulk single crystal of $CrPS_4$ mounted on an 8-circle goniometer with a Pilatus 200K detector at the BL-13, INDUS-II synchrotron radiation source with 10 keV photon energy.

**Raman Measurement:** The Raman spectroscopy measurement on a $CrPS_4$ single crystal was done using a Horiba Jobin Yvon HR-800 single spectrometer in back-scattering configuration. Raman data were taken using a 632.8 nm excitation wavelength, an 1800 g/mm grating and a CCD detector with an overall spectral resolution of ~1 $cm^{-1}$. The power of the laser was kept at ~ 3-4 mW to avoid local heating. Temperature-dependent Raman spectra were taken from 4 K to 300 K using a liquid He flow-type cryostat ST-500 USA, with a temperature stability of 0.1 K. The Raman data on the $CrPS_4/In_2Se_3$ heterostructure were taken using a 532 nm excitation laser attached with Oxford make AFM coupled WITec Alpha 300 spectrometer with a 2400 g/mm grating and a CCD detector. The thickness of the heterostructure was measured using the coupled-AFM system.

**Magnetisation Measurement:** Magnetisation measurement were performed with a Quantum Design SQUID –VSM magnetometer.The measurements were taken from 4 K to 300 K in magnetic field of 100 Oe and 2T field applied perpendicular to the crystallographic *c*-axis of $CrPS_4$.

**TEM sample preparation and imaging:** Cross-sectional and plan-view $In_2Se_3$ flakes were imaged in the HAADF-STEM mode using a Thermo Fisher Scientific TITAN Themis 300 TEM, equipped with a Cs aberration corrector and with a SuperXG quad EDS detector, operated at an accelerating voltage of 300 kV. A low probe current of approximately 20 pA was used to minimize beam-induced damage. The probe convergence angle was set to 24.5 mrad, with a HAADF detector collection angle ranging from 48 to 196 mrad.

For the cross-sectional imaging, the specimen was prepared from $CrPS_4$ crystals using FIB-Helios 5 UX dual-beam system. For plan-view imaging, a few-layer $CrPS_4$ flake was mechanically transferred onto a Cu grid using a dry mechanical transfer method.

**X-Ray Absorption spectroscopy:** XAS studies were carried out at the Cr L-edges at the BL-01 beamline at the INDUS-II synchrotron radiation source in total electron yield mode [[38]]. The edges were recorded at room temperature and at ~100 K using a cold finger liquid $N_2$ cryostat.

**Resistance measurement:** Temperature-dependent linear four probe resistance were measured in cooling cycle from 300 K to 4 K using a Cryogen-Free Measurement System (CFMS, Cryogenic Limited). A constant current of 0.1 μA was applied to the sample using Keithley 2450 Source Meter, and the corresponding voltage was measured using Keithley 2182A nano-voltmeter.

**Photoconductivity measurements:** I-V were recorded from 4K to 300K in a Cryo stage (ST-500) under 632.8 nm excitation laser light of 2 mW at the sample in OFF and ON condition. Current and voltage were applied using a commercial setup in a two-probe configuration using a Keithley 2425 source meter and a Keithley 2002 multimeter.

**Heterostructure preparation:** $CrPS_4$ and $In_2Se_3$ flakes are exfoliated by mechanical scotch tape based exfoliation and used for making $CrPS_4/In_2Se_3$ heterostrucutre via PDMS based dry transfer method.

## 3. Results and discussion

## A. Characterization

$CrPS_4$ crystallizes in a monoclinic structure with space group $C_2$, and lattice parameters $a$ = 10.87 Å, $b$ = 7.25 Å, $c$ = 6.14 Å, and $\alpha = \gamma = 90°$, $\beta = 91.88°$ as reported earlier [29]. VESTA image [39] of side and top views of layered $CrPS_4$ is shown in Figure 1(a). In this compound, the chromium atoms are surrounded by distorted sulphur octahedra, while phosphorus atoms are at the centre of sulphur tetrahedra. The edge-sharing $CrS_6$ octahedral columns along *b*- axis is connected by $PS_4$ tetrahedra, which makes the crystal structure highly anisotropic. This structural anisotropy leads to an anisotropic phonon behaviour probed by angle-resolved polarised Raman spectroscopy (ARPRS) and shown in Figure S1. This behaviour matches well with previous reports [40].

Atomic resolution scanning transmission electron microscopy (STEM) reveals the structural anisotropy in $CrPS_4$. Figure 1(b) and 1(c) show the HAADF-STEM images of the top and side views of $CrPS_4$. The images show a well-defined, ordered structure with no significant structural defects. The structural model of $CrPS_4$ is overlaid on the STEM image and matches very well (shown in the inset in Figure 1(b)), confirming the high crystallinity of $CrPS_4$. Edge-sharing $CrS_6$ octahedra are along the *b*-axis and make the crystal highly anisotropic in the *a-b* plane. The side view clearly demonstrates the layered nature of $CrPS_4$. Energy-dispersive X-ray spectroscopy (EDS) confirms uniform distribution of Cr, P, and S (shown in Supplementary Figure S2).

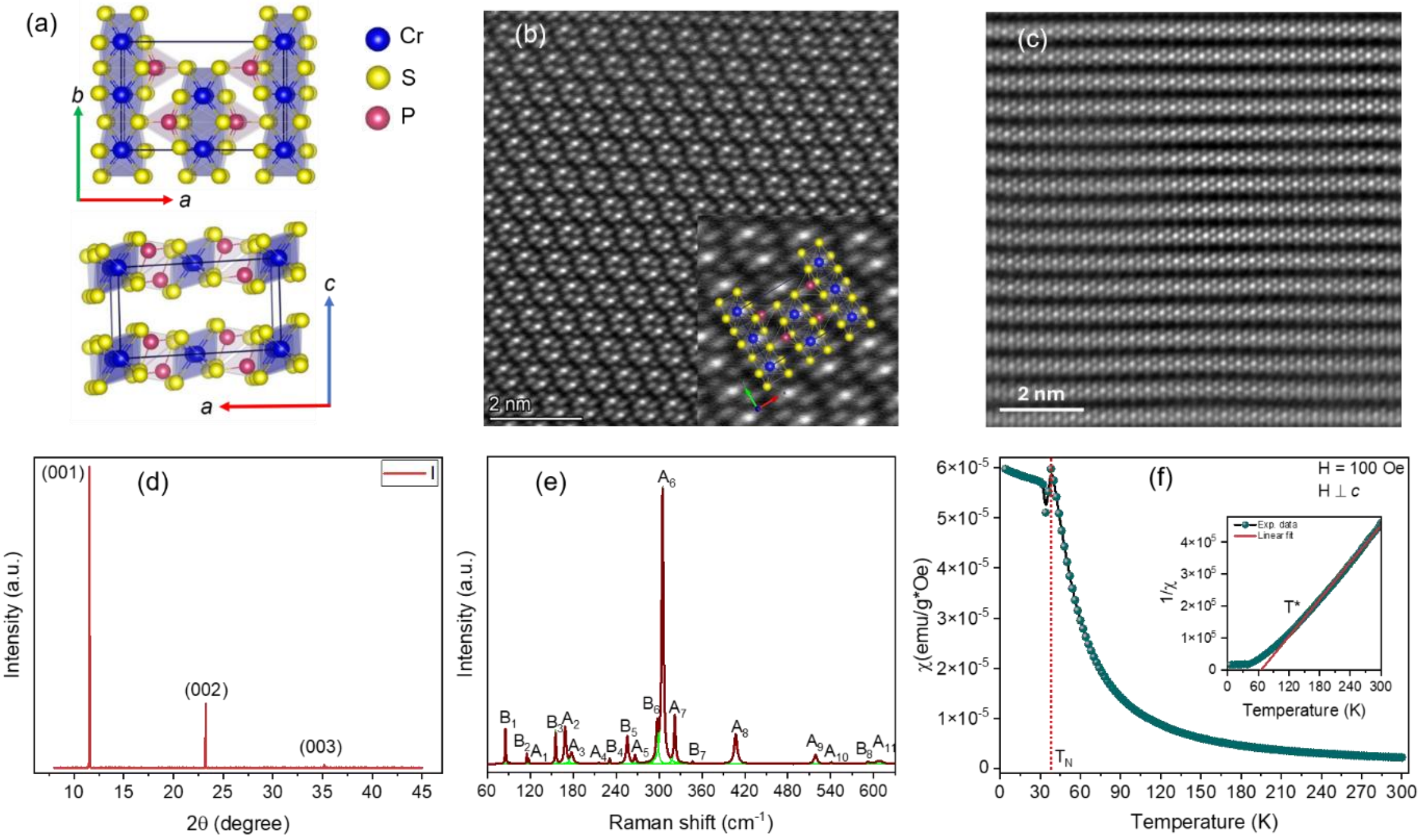

Figure 1: (a) Top view and side view of the $CrPS_4$ crystal structure. (b) Top view and (c) side view of HAADF- STEM image of a single crystal of $CrPS_4$. (d,e) Out-of-plane XRD θ-2θ scan and Raman spectra of the $CrPS_4$ single crystal. (f) Susceptibility χ vs T under 100 Oe external magnetic field applied perpendicular to the *c*-axis measured in zero-field-cooled condition. $T_N$ ~ 40 K marks the Néel temperature. The inset shows the inverse of susceptibility vs. T along with Curie-Weiss fitting (red solid line). T*~120K marks the onset of magnetization deviation from Curie-Weiss behaviour.

XRD θ-2θ scan on a $CrPS_4$ single crystal (Figure 1(d)) shows only the (00C) reflection, confirming the high quality of the grown crystal and its orientation. Figure 1(e) shows the Raman spectrum collected on *a-b* plane. The monoclinic crystal structure gives rise to a large number of well-defined Raman modes. Based on previous report [28] these Raman modes were assigned as A and B modes, as summarised in Table S1.

Magnetic properties were probed using SQUID VSM measurements. Temperature-dependent magnetisation (M) under zero-field-cooled conditions at 100 Oe external field (H) applied perpendicular to the *c* axis is shown in Figure 1(f). The inflection point in $\chi$ (M/H) around 40 K ($T_N$) indicates a magnetic phase transition from the paramagnetic to the antiferromagnetic phase [24]. The inset of Figure 1(f) shows $1/\chi$ vs T along with a Curie-Weiss fit (red solid line). The fit deviates from experimentally observed $1/\chi$ around T*~120 K indicating setting in of short-range spin order. Temperature-dependent magnetisation was also carried out under a higher field of ~2 T, which is shown in the Supplementary Figure S3. At that field, the cusp at $T_N$ was suppressed.

## B. Temperature-dependent Raman spectrum

To probe the temperature evolution of the lattice dynamics, we performed a temperature-dependent Raman spectroscopy study (4- 300 K). The $CrPS_4$ unit cell contains 12 atoms and thus has 36 phonon modes (17A+19B), of which three are acoustic A+2B, and 33 are optical phonons [[41],[42]]. A total of 19 Raman-active modes were observed in the present study up to 650 $cm^{-1}$, which match well with previous studies [[43]]. The experimentally observed frequencies of the phonon modes are obtained by fitting the spectra using Lorentzian functions and are presented in Table S1 along with atomic vibrations taken from the literature [28]. In this measured temperature window, no extra peak was observed, suggesting the absence of a structural transition.

Raman spectra collected at various temperatures are stacked in Figure S4, and to see the temperature evolution, magnified views are also shown in Figure S4(b,c and d). We observed softening of the Raman modes with increasing temperature, which is attributed to phonon anharmonicity [[44]]. The Raman mode position (wavenumber) and width (FWHM) of selected Raman modes as a function of temperature are presented in Figures 2 and 3, respectively. [[45],[46]] and the remaining Raman mode frequencies with temperature are presented in Figure S5. Among them, $B_1$, $B_3$, $B_4$ and $A_3$ modes are related to mainly $CrS_6$ octahedral vibrations, while $A_8$ and $A_9$ are related to $PS_4$ tetrahedra vibrations (see Table S1). It is worth noting that Raman modes related to $CrS_6$ octahedra showed deviation from anharmonic behaviour at T*, whereas modes related to $PS_4$ tetrahedra ($A_8$, $A_9$) showed deviation at $T_N$. This suggests an isostructural distortion in $CrS_6$ octahedra around T*.

We adopted the three-phonon decay model of Balkanski *et al* [45], for fitting the anharmonic behaviour. According to this model, the temperature variation of the Raman mode frequency can be expressed as

$$\omega_{anh} = \omega(0) - A\left[1 + \frac{2}{\exp\left(\frac{\hbar\omega}{2kT}\right)-1}\right] \quad [1]$$

and the linewidth can be given as

$$\Gamma = \Gamma_0\left[1 + \frac{2}{\exp\left(\frac{\hbar\omega}{2kT}\right)-1}\right] \quad [2]$$

The temperature evolution of the Raman mode position and width is fitted using equations 1 and 2, respectively (solid red line in Figures 2,3 and Figure S5). The fit matches well up to 130 K, while a significant deviation from the anharmonic behaviour is observed below ~130 K, along with a sudden change at $T_N$ (40 K). The observed anomalies at $T_N$ is similar to the earlier reports and signify the presence of strong phonon modulation due to the setting in of long-range magnetic ordering [28]. We computed the value of spin-phonon coupling for each Raman mode.

The changes in Phonon frequency as a function of temperature are mainly due to anharmonic terms, in addition to changes due to spin ordering [[47],[48]].

$$\Delta\omega_{Total} = \Delta\omega_{anh} + \Delta\omega_{spin}$$

So

$$\Delta\omega_{spin} = \Delta\omega_{Total} - \Delta\omega_{anh}$$

Change in phonon frequency due to magnetic ordering ($\Delta\omega_{spin} = \Delta\omega_{Total} - \Delta\omega_{anh}$) can be represented by the spin-spin correlation function $\langle S_i \cdot S_j \rangle$

$$\Delta\omega_{spin} = \lambda\langle S_i . S_j \rangle$$

For antiferromagnetic correlations, the change in frequency can be described as [49]

$$\Delta\omega = -\lambda S^2 \phi(T) \quad [3]$$

Where $\phi(T) = \left(1 - \frac{T}{T_N}\right)^{\gamma}$ [50]

$\phi(T)$ is an order parameter, S=3/2 for the unit cell of $CrPS_4$, and λ is the spin-phonon coupling constant ($\lambda = \frac{1}{2\mu\omega}\frac{\partial^2 J}{\partial u^2}$) [51], which depends on the effective mass of an atom μ, the frequency of phonon modes $\omega$, and how the exchange interaction $J$ is modified with the vibrational displacement, $u$. We calculated the spin-phonon coupling constant and listed it in Table 1 for each Raman mode; the fitting by equation 3 is shown in supplementary Figure S6.

The linewidth of Raman modes related to $CrS_6$ octahedral vibrations also showed deviations from anharmonic behaviour (Figure 3) at T*. In contrast, the Raman modes corresponding to $PS_4$ tetrahedra deviate only in the vicinity of $T_N$, confirming that the lattice modifications are related to $CrS_6$ octahedral distortion. Additionally, to probe the interplay of lattice and electronic properties, we performed electrical transport and photoconductivity studies.

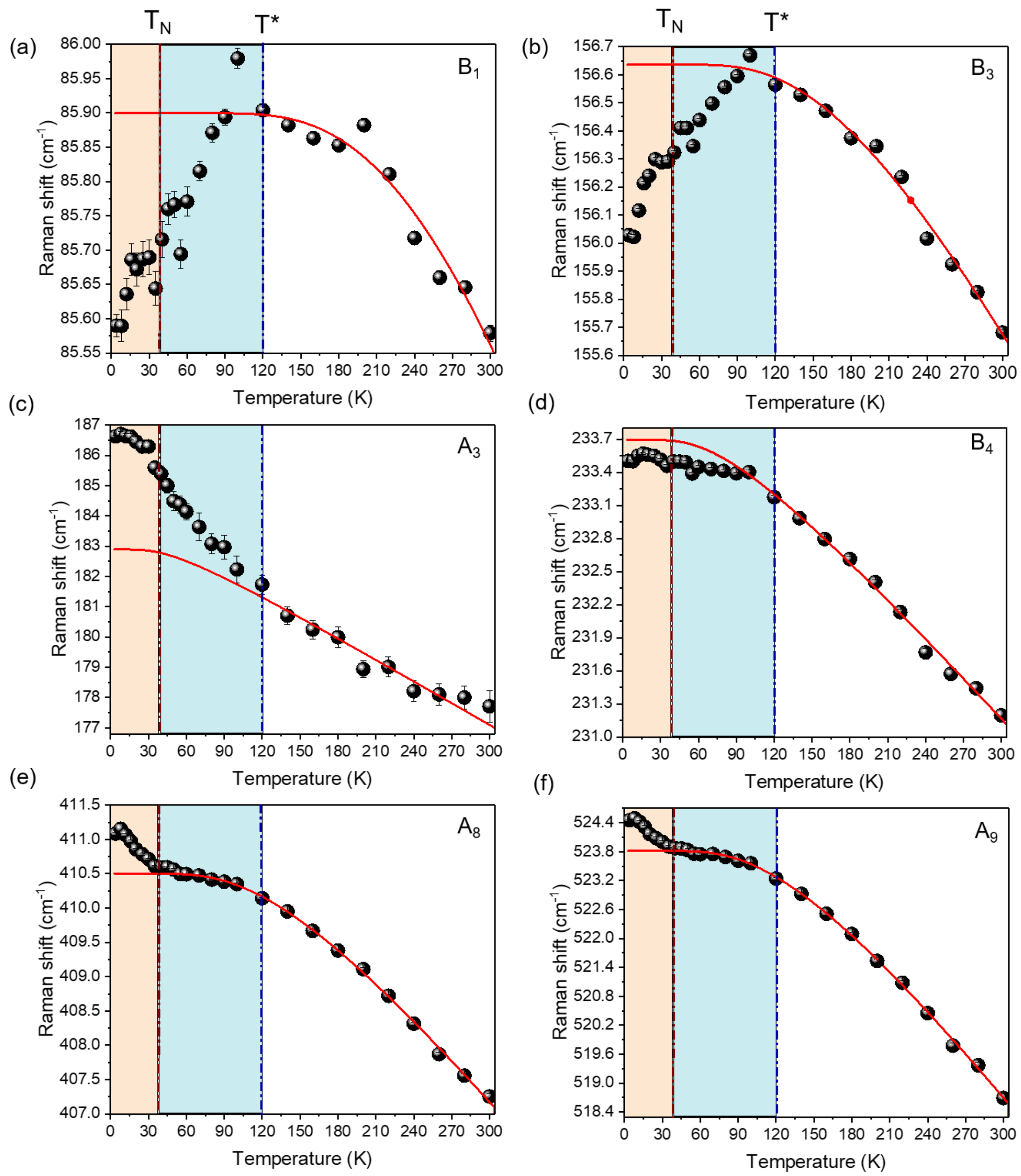

Figure 2: (a)-(f) Temperature-dependent Raman shift of selected Raman modes. $B_N$ and $A_N$ are Raman modes assigned in Table S1. Black dots are experimental values of the Raman shift, while the solid red lines show anharmonic fitting. The shaded region highlights two different regimes: the blue region is marked from T* to $T_N \sim 40$ K, and the orange Region is marked below $T_N$ (long-range magnetically ordered phase).

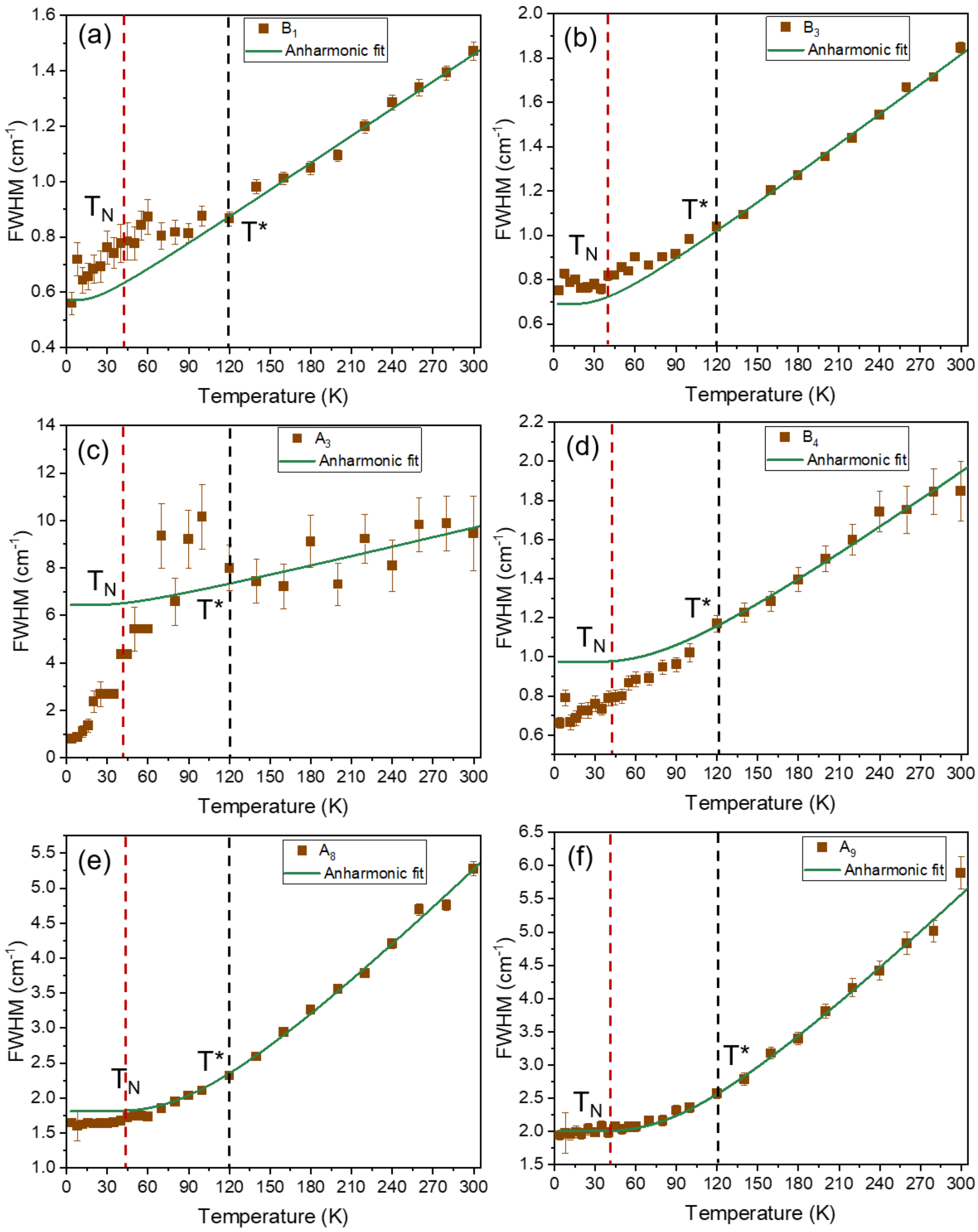

Figure 3: (a)-(f) Temperature-dependent linewidth of selected Raman modes. Brown boxes are experimental values of the linewidth of the Raman modes, while the solid green lines represent anharmonic fitting using a three-phonon decay process

**Table 1:** Experimentally obtained spin-Phonon coupling value

.

| Raman mode Frequency | λ (Spin-phonon coupling strength) |
|---|---|
| $B_1$ (84.8 $cm^{-1}$) | 0.105 |
| $B_2$ (115 $cm^{-1}$) | 0.064 |
| $A_1$ (117.7 $cm^{-1}$) | 0.045 |
| $B_3$ (155 $cm^{-1}$) | 0.25 |
| $A_3$ (177.4 $cm^{-1}$) | -1.375 |
| $A_4$ (215.3 $cm^{-1}$) | 0.11 |
| $B_4$ (231 $cm^{-1}$) | 0.06 |
| $A_6$ (304.5 $cm^{-1}$) | 0.49 |
| $A_8$ (407 $cm^{-1}$) | -0.26 |
| $A_9$ (519 $cm^{-1}$) | -0.32 |
| $A_{10}$ (540.7 $cm^{-1}$) | 0.07 |
| $A_{11}$ (607.6 $cm^{-1}$) | 0.75 |

## C. Temperature-dependent Resistance

To further probe the effect of the lattice distortion observed at T* in Raman spectroscopy on electronic properties, we measured the electrical transport and photoconductivity as a function of temperature. Figure 4(a) shows the temperature-dependent resistance along the *b*-axis collected while cooling from 300 K to 4 K under an applied current of 0.1 μA. The resistance showed an increase with decreasing temperature, indicating semiconducting characteristics of $CrPS_4$ [52]. However, it shows two anomalies: one at T*, below which it showed a sudden rise and the second one at $T_N$, below which the resistance becomes temperature-independent. Further, photoconductivity, i.e. current vs. voltage (I-V) measurements as a function of temperature under a 632.8 nm laser in OFF and ON conditions (Figure S7(a) & (b)) were carried out. From the I-V curves, we extracted the value of current at -10 V at various temperatures under laser OFF and ON protocols, and the difference in these obtained values, i.e. ($I_{OFF}$-$I_{ON}$ at V=-10 V) as a function of T, is plotted in Figure 4 b, which denotes photoconductivity as a function of T. The photoconductivity shows anomalies across T* and $T_N$ (details are shown in Supplementary Figure S7(c) &(d)).

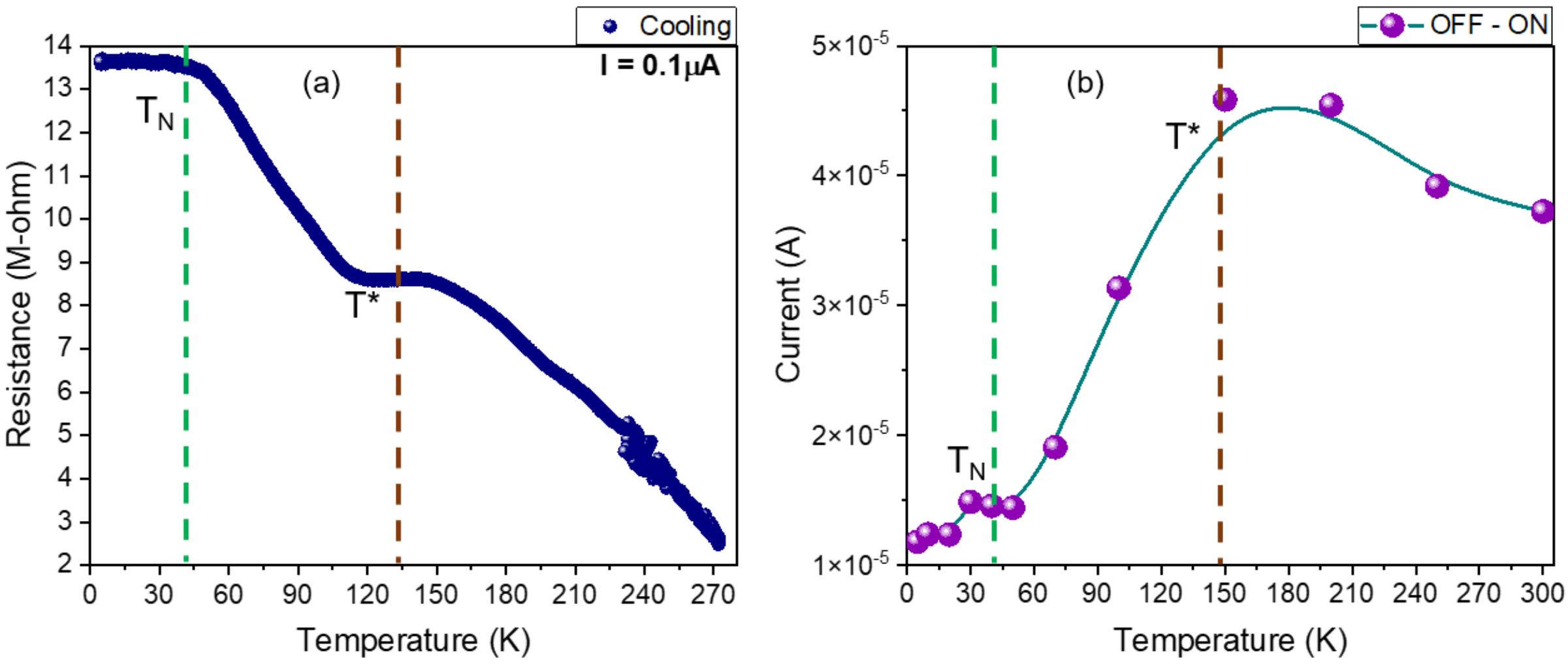

Figure 4: (a) Temperature-dependent Resistance of $CrPS_4$. (b) Photoconductivity as a function of temperature at -10V.

## D. Temperature-dependent Photoluminescence

To further investigate the changes in electronic structure, we measured photoluminescence (PL) as a function of temperature, which is shown in Figure 5. The photogenerated excitons are mostly dominated by $Cr^{3+}$ d-d transitions [53,54] and therefore, PL characteristics mainly arise from the $Cr^{3+}$ electronic state transitions only [31]. The observed PL arises from the $^2E$ to $^4A_2$ transition [31,55]. $^4A_2$ is defined as all three electrons of $Cr^{3+}$ in the $t_{2g}$ ($t_{2g}^{\uparrow\uparrow\uparrow}$) orbital with parallel spin, while the $^2E$ represents the excited state, where one electron's spin is flipped ($t_{2g}^{\uparrow\uparrow\downarrow}$). Under the spin conservation rule, this transition is forbidden; however, due to hybridisation of Cr 3d-orbitals with the ligand (S here), this selection rule is relaxed. The PL is also observed due to a transition from $^4T_2$ to $^4A_2$, which has an energy of 1.1 eV, which is beyond the measurement range of the set-up used in the present studies [31]. We measured the PL from 1.24 eV upwards, and the PL at selected temperatures are shown in Figure 5(b): 40 K, 5(c): 120 K and 5(e): 300 K, where the PL exhibited distinct changes.

PL spectra were fitted using a Gaussian function to obtain the peak parameters. At 300 K, electrons de-excite dominantly via non-radiative transition and therefore the PL spectrum is very weak. The PL spectra at room temperature consist of two broad peaks, and several sub-peaks or overtones begin to appear as the temperature decreases; their intensity is significantly enhanced below 150K. However, below $T_N$, the PL spectra change drastically and become sharp [see Figure S11, Supplementary Information]. The observed overtone features with the lowering temperature indicate strong vibronic coupling, indicating enhanced electron-phonon coupling strength. A schematic representation of the vibronic optical transition is shown in Figure 5(d) (discussed in detail in the discussion section). The electron-phonon coupling strength is obtained in the form of the Huang–Rhys factor S [56]. We extracted S from the Franck-Condon relation [57] $I_n = I_{ZPL} e^{-S} \frac{S^n}{n!}$, where $I_n$ is the intensity of the n-vibronic couplet transition, $I_{ZPL}$ is the intensity of the zero-phonon line. We calculated S from the relative intensities of the first-order phonon replica and the zero-phonon transition, i.e. $S = {I_{0-1}}/{I_{0-0}}$ at each temperature and presented it in Figure 5(f). Interestingly, it also showed a sudden rise at~T*.

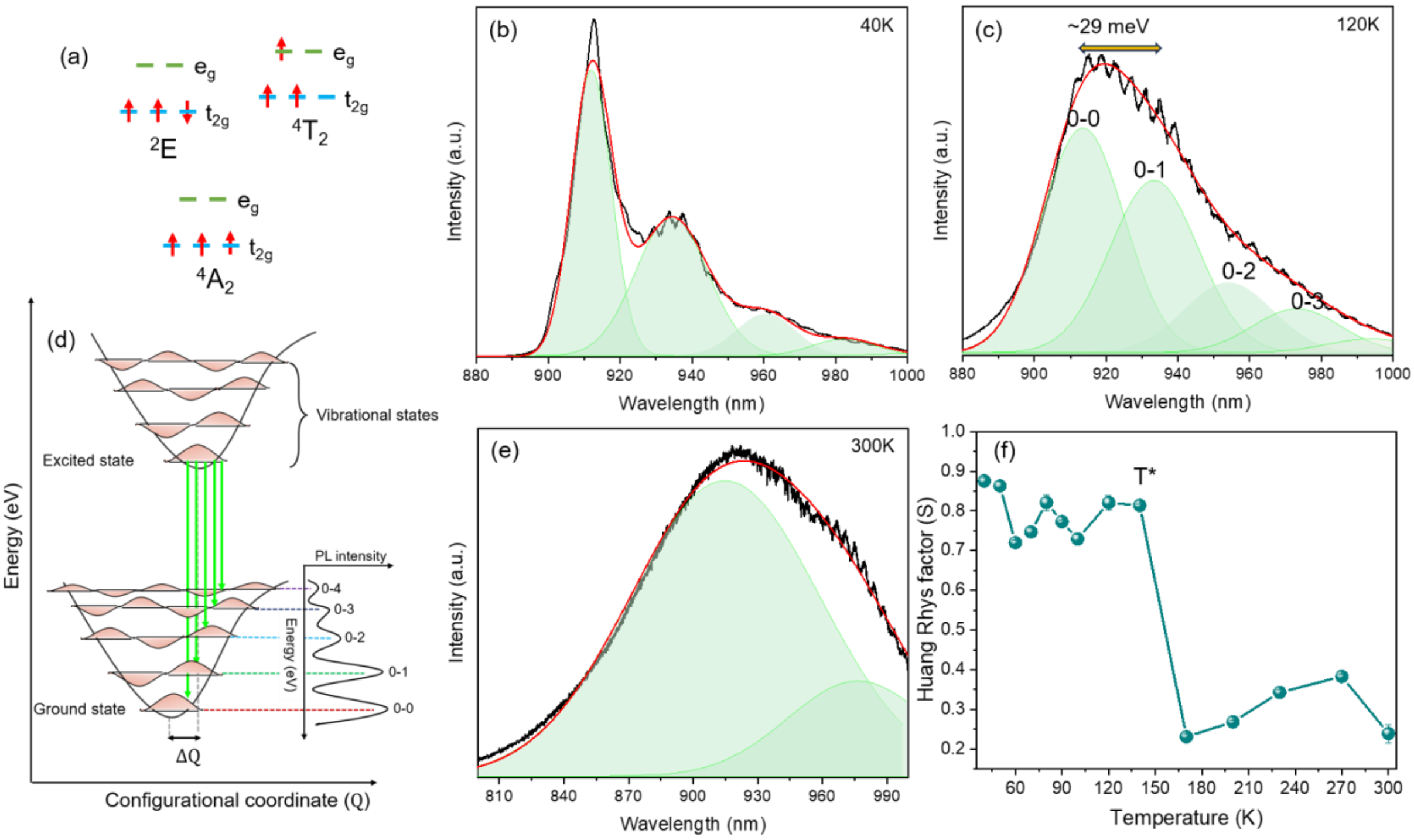

Figure 5: (a) Electronic states of $Cr^{3+}$, showing $^4A_2$, $^2E$ and $^4T_2$ spin-state configurations. (b), (c) and (e) PL spectra at 40 K, 120 K, and 300 K, respectively. Low-temperature PL show multiple subpeaks, involving the zero-phonon line (0-0) and their corresponding phonon replicas like 0-1, 0-2,... having a separation of ~ 29 meV (d) Energy vs configuration coordinate diagram of the 3d $Cr^{3+}$ showing multiple phonon-assisted optical transitions. (f) Temperature-dependent Huang-Rhys factor S determined from the relative intensity of the zero-phonon line (0-0) and first phonon replica (0-1).

## 4. A. Heterostructure of $CrPS_4/In_2Se_3$

To exploit the strong spin-phonon/spin lattice coupling observed in $CrPS_4$ around T*, a heterostructure is attempted with $In_2Se_3$. $In_2Se_3$ is a well-known 2D ferroelectric material with above room temperature ferroelectric transition temperature. Thus, the heterostructure provides a fertile playground for studying the multiferroic properties. The choice of this combination has several advantages, first, a two-dimensional material can be easily stamped onto another to create a desired heterostructure and functionality; second, the interfacial effect of strong spin-phonon coupling and lattice distortion observed in $CrPS_4$ on a ferroelectric material $In_2Se_3$ can be tested.

$CrPS_4$ bulk crystals were mechanically exfoliated by scotch tape and transferred onto a $Si/SiO_2$ substrate. Then $In_2Se_3$ were transferred onto $CrPS_4$ by using a 2D transfer system (details are in experimental section). An optical image of the heterostructure is shown in Figure 6 (a). The isolated $CrPS_4$ layer is marked with a green star, while the isolated $In_2Se_3$ layer is marked with a black star, and the overlapping region is marked with a red star. Room-temperature Raman spectra were taken on isolated $In_2Se_3$ (region marked by a black star), isolated $CrPS_4$ (green star) and on the heterostructure (red star) using a 532 nm excitation source and are shown in Figure 6 (b). As reported previously, In $In_2Se_3$, modes at 26.4 $cm^{-1}$ and 89.4 $cm^{-1}$ are identified as E modes, while the high-intensity 104 $cm^{-1}$ Raman mode is of A character, and higher-frequency modes at 180, 186 and 197 $cm^{-1}$ correspond to A(LO), E and A(TO) modes, respectively [[58]]. The green spectra corresponding to isolated $CrPS_4$ also matched well with the bulk $CrPS_4$ spectra shown in Figure 1. Red spectra recorded on the heterostructure region showed the Raman modes of both layers, $In_2Se_3$ and $CrPS_4$. In this spectrum, for clarity, we marked the Raman modes of $In_2Se_3$ with black stars and those of $CrPS_4$ with green stars. A clear shift in the peak position of $In_2Se_3$ in the isolated and heterostructure regions is observed; for example, in the heterostructure, the E mode peak is observed at 26.7 $cm^{-1}$ compared to 26.4 $cm^{-1}$ in the isolated layer. Similarly, the A mode position is 103.9 $cm^{-1}$ in the heterostructure; while it is observed at 103.6 $cm^{-1}$ in the isolated layer. This establishes the coupling between the two layers in the heterostructure. Figure 6(c) shows the AFM mapping of this flake. The thickness of the two isolated regions has been determined from the AFM and is found to be ~42 nm for $CrPS_4$, and ~ 96 nm for $In_2Se_3$. Raman mapping was also done to map the spatial distribution of $In_2Se_3$ and $CrPS_4$ in the heterostructure, which matched the optical image (see Supplementary Figure S9).

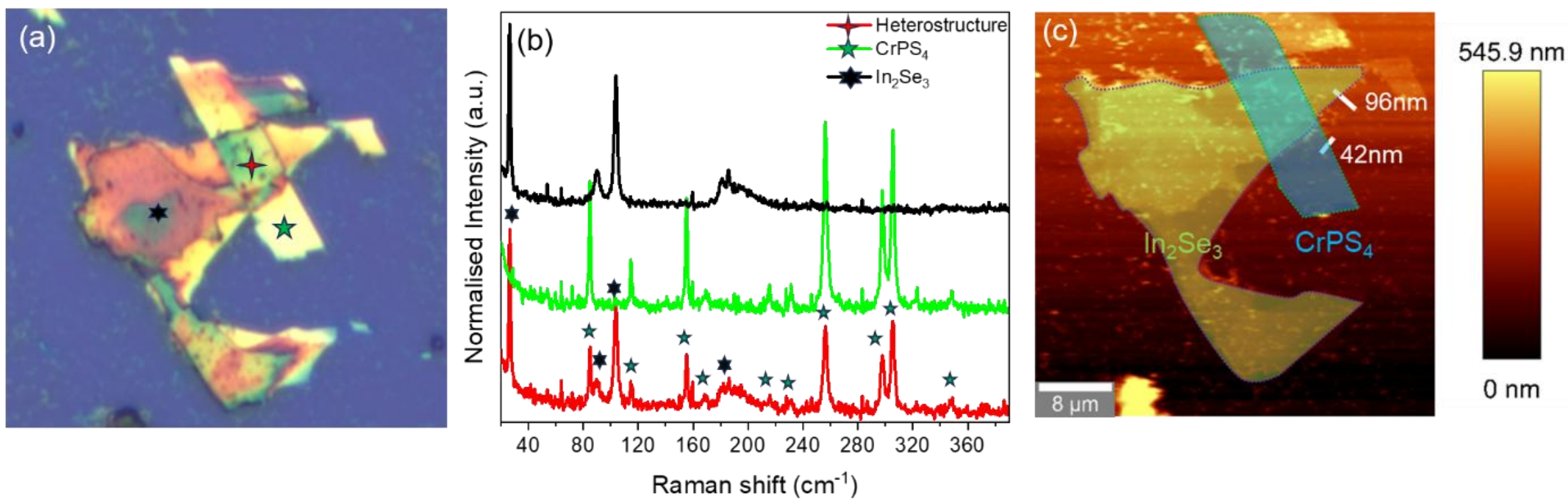

Figure 6: heterostructure of $CrPS_4/In_2Se_3$. (a) Optical image of heterostructure. Red star is in the region of the heterostructure of $CrPS_4$/ $In_2Se_3$, Green star shows isolated $CrPS_4$, and Black star shows isolated $In_2Se_3$. Raman spectra were collected at the three points (b). Black is the Raman spectrum of isolated $In_2Se_3$, Green is the Raman spectrum of isolated $CrPS_4$, and Red is the Raman spectrum of the heterostructure showing both Raman modes of $CrPS_4$ and $In_2Se_3$. (c) AFM mapping of the heterostructure for obtaining thickness of layers: $In_2Se_3$ ~ 96nm, and $CrPS_4$ ~42 nm.

## B. Temperature-dependent Raman spectra of the heterostructure

To explore the proximity effect of magnetic ordering of $CrPS_4$ on $In_2Se_3$, we carried out temperature-dependent Raman spectra on this heterostructure from 10 K to 300 K. Raman spectra at different temperatures were recorded on the heterostructure and isolated $In_2Se_3$ (shown in Figure S10) and were fitted using Lorentzian functions. The T-dependence of the frequency of Raman modes of isolated $In_2Se_3$ and the heterostructure is plotted in Figure 7. The T-dependent frequency of the Raman modes is fitted using the anharmonic model described before (equation 1), which is shown by a pink solid line, while the experimental peak positions are shown by blue sphere (Figure 7 (a-c)). Similarly, the peak positions of the Raman modes of $In_2Se_3$ in the heterostructure is deduced and presented in Figure 7(d)-(f); experimental peak positions are shown by brown dots and anharmonic fitting by a green solid line. In Figures (a)-(c), no phonon anomaly was observed in the peak position of the isolated $In_2Se_3$, and the behaviour is well represented by the anharmonic model.

However, in the heterostructure, the T-dependence of Raman mode frequencies of $In_2Se_3$ deviates from the anharmonic fit at $T_N$ ~ 40 K of $CrPS_4$. Interestingly, the E mode (~90 $cm^{-1}$) of $In_2Se_3$ (Figure 7(e)) showed marginal deviation from anharmonic behaviour from T*. The deviation from anharmonic behaviour is more prominently seen in the T-dependence of line width (FWHM) of these modes shown in Figure 8. As expected, no Anomaly in the phonon linewidth is observed in isolated $In_2Se_3$, whereas in the heterostructure, deviations from anharmonicity were observed from T*. This confirms that the phonon anomalies in the heterostructure $In_2Se_3$ are due to the proximity effect induced by $CrPS_4$ magnetic anomalies discussed earlier. Therefore, it demonstrates the modification of the lattice dynamics of a ferroelectric layer by spin correlation of a magnetic layer via interlayer coupling. This is an important result, as the spin–phonon coupling is strongly manifested in the Raman shifts of $In_2Se_3$ due to the magnetic ordering of $CrPS_4$, suggesting its potential for establishing multifunctional correlation.

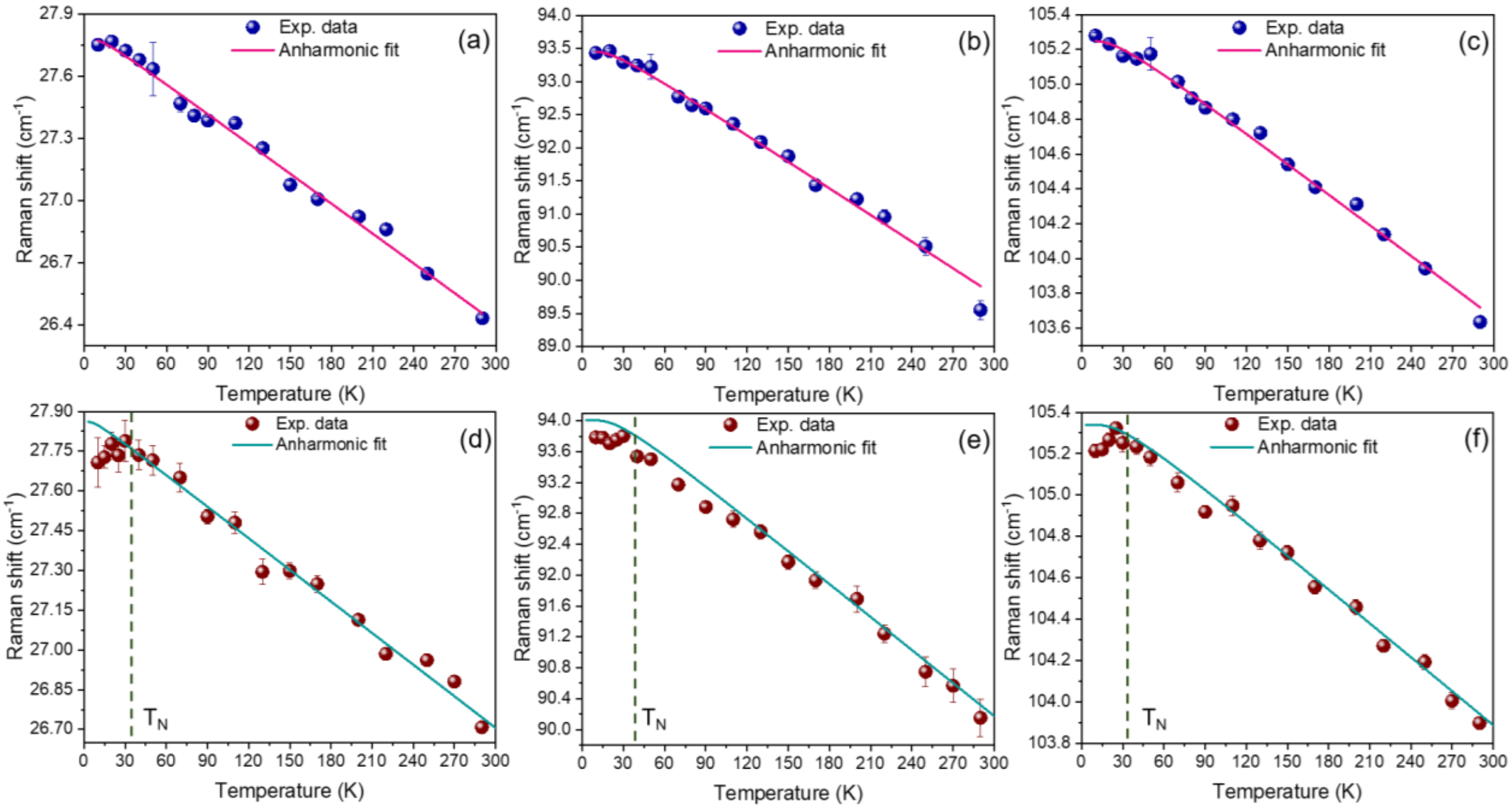

Figure 7: T- evolution of phonon modes of $In_2Se_3$: (a), (b) and (c) Raman modes of isolated $In_2Se_3$. Blue squares represent experimental data, and the solid red line represents the fitted anharmonic equation. The anharmonic equation over the whole temperature range fits the $In_2Se_3$ Raman modes. (d), (e) and (f) represent the same Raman modes of $In_2Se_3$ in a heterostructure. Brown dots represent experimental data, and the green curve is fitted with anharmonicity. Below 40 K, these Raman modes deviate from anharmonicity due to the influence of magnetic ordering in $CrPS_4$.

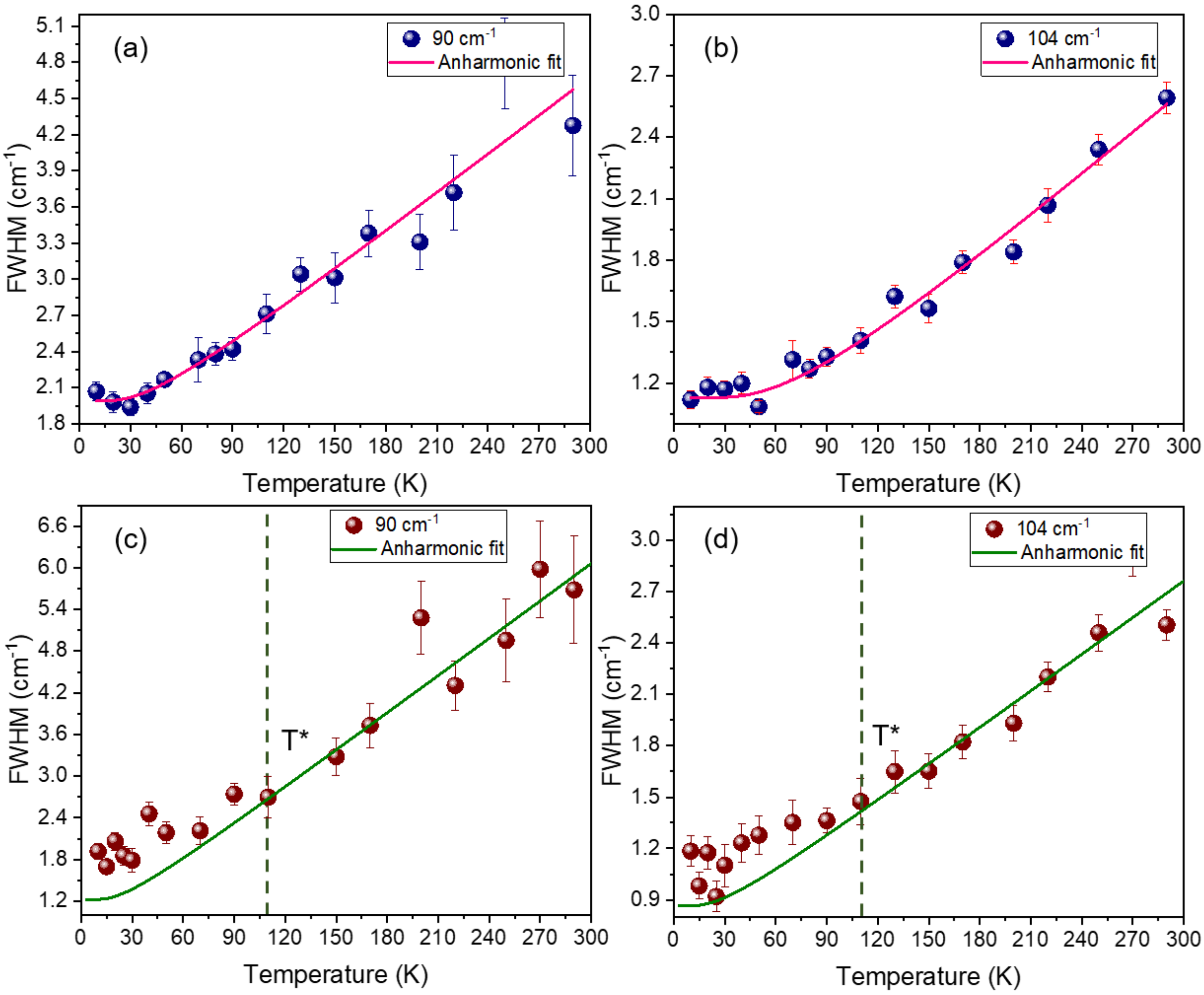

Figure 8: T evolution of linewidth of Raman modes of isolated $In_2Se_3$ (a) and (b) in the heterostructure (c) and (d). Clearly, linewidths deviate from the anharmonic equation at 120 K.

## 5. Discussions

In the present study, we observe the onset of spin–phonon coupling in $CrPS_4$ at T* ~120 K, well above $T_N$. By fabricating a $CrPS_4/In_2Se_3$ heterostructure, we further demonstrate that the lattice modification associated with this spin–phonon coupling can be transferred to the adjacent $In_2Se_3$ layer through interfacial coupling. We will discuss now the possible root cause for the enhanced temperature T*.

Former temperature-dependent X-ray diffraction studies illustrated lattice modification in the *b*-lattice parameter, showing minima around ~130 K. It is worth noting that the *b*- lattice parameter depends strongly on the Cr-Cr bond length, suggesting temperature-dependent modification of $CrS_6$ octahedra [29]. In order to probe the modification in $CrS_6$ octahedra a local probe is required and Raman studies are thus an ideal option.

Our temperature-dependent Raman study confirmed this isostructural modification. The Raman peak position and FWHM show a clear change at T*, shown in Figures 2 and 3, respectively. Temperature-dependent Raman measurements further confirm the absence of any new Raman modes in the whole temperature window, confirming no symmetry changes (the lattice parameter *b*-anomaly only gives an isostructural modification, not a phase transition). Importantly, only phonon modes involving $CrS_6$ octahedral vibration showed strong changes at T*, which directly indicates strong spin-lattice coupling.

The largest value of spin-phonon coupling is also observed for the $CrS_6$ octahedral breathing vibration ($A_3$) Raman mode, because the octahedral breathing vibration directly modulate the Cr-S bonds and hence largely effect on the in-plane exchange interaction. These observations unequivocally suggest that short-range magnetic correlations persist well above $T_N$ in $CrPS_4$ and the distortion in $CrS_6$ octahedra is strongly coupled to these correlations. This material is highly anisotropic, it shows a ferromagnetic spin order in the *a-b* plane, while these layers are antiferromagnetically coupled along the *c*-axis. It is highly likely that the in-plane ferromagnetic exchange is stronger than the out-of-plane antiferromagnetic exchange and, thus, persists well above $T_N$ where it provides short-range magnetic correlations. These results are in consonance with recent magnetic pair distribution function analysis, which showed that ferromagnetic intrachain correlations persist far above $T_N$ [[59]]. This is also reflected in our electrical, optical, as well as magnetic measurements. In particular, our magnetic data for $1/\chi$ vs T revealed deviations from the Curie-Weiss fit below T* ~120 K (Figure 1(f) and Figure S(3)).

In our electrical and optoelectronic measurements (Figure 4(a) and 4(b)), the temperature-dependent resistance also showed anomalous behaviour around T*. The octahedral distortion changes the Cr-S-Cr bond length and angle, which modify the electron hopping strength and hence the electron transport. It is to be noted that the long-range spin ordering reduces spin dependent scattering, which is reflected in a decrease in the slope of resistance as a function of temperature below $T_N$, making it almost T-independent. The photoconductivity with temperature also shows a sudden drop around T*, indicating a modification of the electronic structural environment due to octahedral distortion.

In order to confirm the modification in electronic structure due to changes in $CrS_6$ octahedra distortion X-ray absorption spectra are collected at the Cr L-edges at room temperature and at low temperature (~100 K) (shown in supplemental Figure S8). Subtle but observable modification in the spectral line shape and broadening at low temperature (~T*) further reveals the modification in the local ligand field environment of the Cr atom. These modifications can arise due to core hole potential, core hole induced charge transfer effect, crystal field, core hole–valence hole exchange and multipole interaction [[60],[61]]. Nonetheless, octahedral distortion effectively modifies those parameters and relate to the observed change in XAS spectra at low temperature.

The changes in Photoluminescence spectra and enhanced vibronic progression below T* can also be understood in terms of modification in $CrS_6$ octahedral distortion. The average vibronic spacing is found to be 29 meV which indicates that the octahedral shear vibrational mode (Raman mode $B_4$) with wavenumber of 234 $cm^{-1}$ is responsible for the vibronic progression (Figure 2(d)). Figure 5(f) shows a sudden enhancement in Huang-Rhys factor S. The Huang–Rhys factor represents electron-phonon coupling strength and is defined as $S = K(\Delta Q)^2 / 2\hbar\omega$ where $(\Delta Q = Q_e - Q_g)$ [[62]], where K is the effective force constant, $\hbar\omega$ is the phonon energy, $\Delta Q$ is the difference between the configurational coordinate of the excited electronic state $Q_e$ and the ground state $Q_g$. Earlier optical absorption studies showed a similar enhancement of S around T* [30]. In $CrPS_4$, the PL originates from the d-d transition of $Cr^{3+}$, where the excited state is $^2E$ and the ground state is $^4A_2$, see Figure 5(a). As the $CrS_6$ octahedral distortion occurs at T*, it should be reflected in PL of $CrPS_4$. In $CrPS_4$, the $Cr^{3+}$ ions are coordinated with six sulphur ions; therefore, the $\Delta Q = \sqrt{6m}\Delta r$ where $\Delta r$ is the change in the Cr-S effective bond length. Around T*, the octahedral modification in $CrS_6$ leads to changes in the effective Cr-S bond length ( $\Delta r$ ) which leads to changes in the configurational coordinate ($increase\ in\ \Delta Q$) of the ground state and excited electronic state shown in Figure 5(d), this leads to coupling of more vibrational states with the electronic state through vibronic progression, as reflected in PL spectra below 120 K compared to 300 K.

# 6. Conclusion

In summary, using various experimental probes, we establish robust multifunctional correlations in the van der Waals antiferromagnet $CrPS_4$ in the form of spin–lattice–electron coupling that persists far above the Néel ordering temperature and is driven by short-range magnetic correlations up to T*. Phonon mode analysis demonstrates that $CrS_6$ octahedral distortions are strongly coupled to these short-range magnetic correlations. Among all active Raman modes, the $A_3$ mode associated with $CrS_6$ octahedral breathing exhibits the largest spin-phonon coupling constant. These structural modifications and the sustained short-range magnetic order below T* macroscopically manifest in the material's magnetization, electrical transport, and optoelectronic profiles. Notably, the optical response couples strongly with both the short-range magnetic order and the octahedral distortion, causing enhanced vibronic progression in the photoluminescence spectra. The energy spacing between the zero-phonon line and its subsequent replicas is ~29 meV, confirming that the octahedral shear vibrational mode ($B_4$ Raman mode) primarily mediates this vibronic progression.

Further, we demonstrated proximity-induced multifunctional correlation between two layers in a $CrPS_4/In_2Se_3$ heterostructure. This is one of the unique studies where the effect of even local lattice distortion in one layer do not remain isolated and affects the adjacent layer through interfacial effects. This work highlights a promising strategy for engineering artificial 2D multifunctional materials tailored for next-generation device applications.

## 7. Acknowledgement

The authors acknowledge Kranti Kumar and Vikas Singh for magnetisation measurements. Dinesh Kumar Shukla and Hemant Singh for XRD measurements and discussions. Rakesh Shah, Akshaya A and Amit Verma for temperature-dependent XAS. The authors acknowledge the National Nanofabrication Centre (NNFC), Micro Nano Characterization Facility (MNCF) and Advanced Facility for Microscopy and Microanalysis (AFMM) facilities of the Indian Institute of Science (IISc), Bengaluru for device fabrication, transport and electron microscopy studies. BKD acknowledge IOE fellowship for financial supports. M.H. acknowledges the support from the Leona M. and Harry B. Helmsley Charitable Trust grant No. 2018PG-ISL006, and from the Israel Science Foundation grant No. 687/22.

## Supplementary information for

# Interplay of spin-lattice and electronic coupling far above Neel ordering in 2D antiferromagnetic $CrPS_4$ and its interface manifestation

Divya Jangra[1], Binoy Krishna De[2]*, Mukesh Kumar Dasoundhi[3], Sourav Chowdhury[4], Pragati Sharma[1], Kartick Biswas[2], Arup Basak[2], Dheeraj Kumar Gupta[1], Koushik chakraborty[1], Praveen Kumar Velpula[1], Pavan Nukala[2], U. Chandni[5], Arvind Kumar yogi[1], Mukul Gupta[1], Markus Hücker[3], Vasant G. Sathe[1]*

[1]UGC-DAE Consortium for Scientific Research, D.A. University Campus, Khandwa Road, Indore-452001, India

[2]Centre for Nano Science and Engineering, Indian Institute of Science, Bangalore-560012, India

[3]Department of Condensed Matter Physics, Weizmann Institute of Science, Rehovot, Israel

[4]Deutsches Elektronen-Synchrotron DESY, Notkestrasse 85, 22607 Hamburg, Germany

[5]Department of Instrumentation and Applied Physics, Indian Institute of Science, Bangalore-560012, India

## 1.0 Polarized Raman spectra

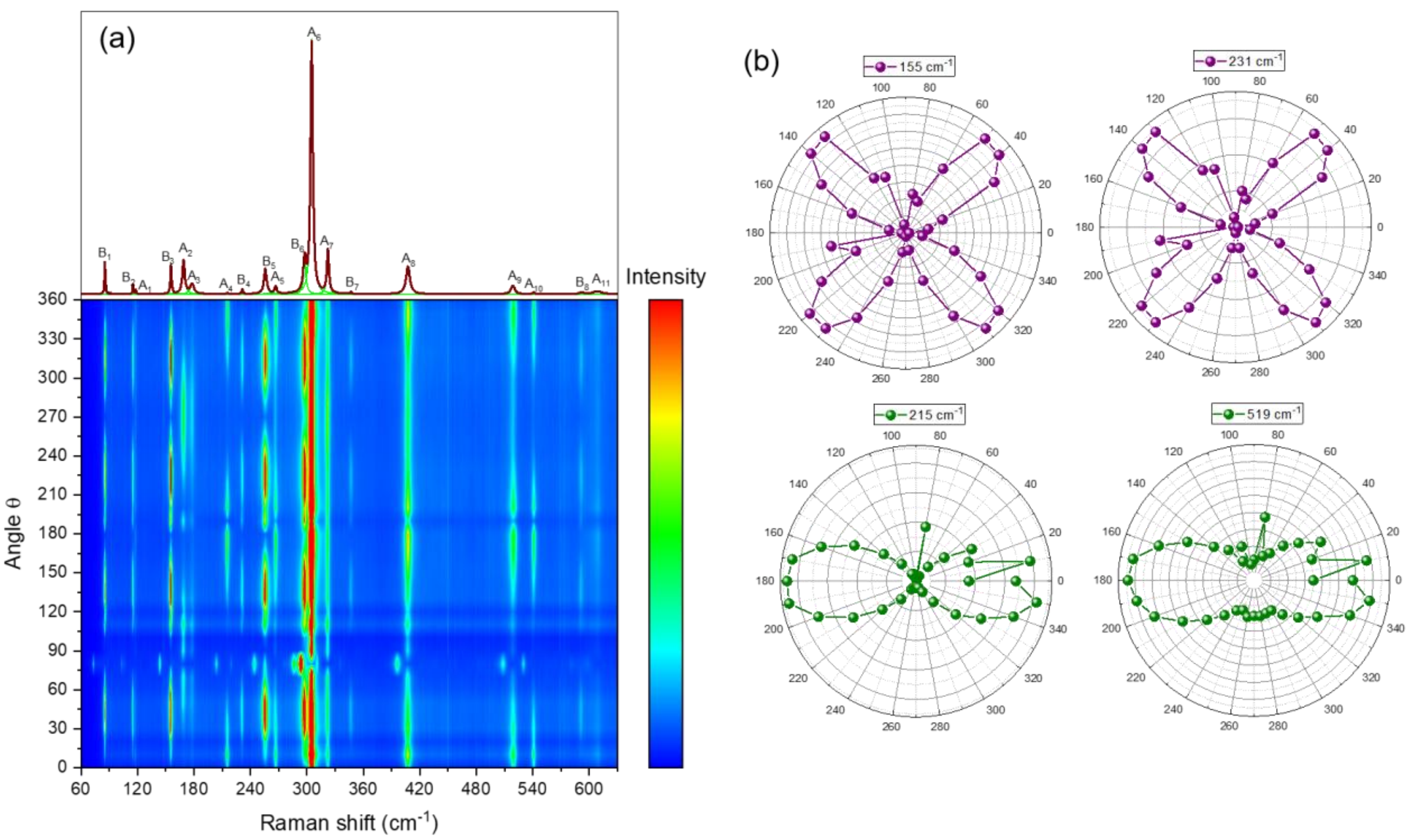

**Figure S1: Angle-Resolved Polarised Raman spectra of single-crystal $CrPS_4$:** The spectra were recorded at room temperature by rotating the sample mounted on a rotator at 10° intervals, keeping the polariser and analyser fixed. (a) Contour plot of Intensity with respect to crystallographic orientation θ rotated from 0° to 360°. At the top of the contour plot, a typical $CrPS_4$ Raman spectrum is shown. Raman modes are labelled according to their symmetry, A or B, and listed in Table-S1. (b) Shows the angular plots of some selected B- and A-symmetry Raman modes' intensity as a function of rotation angle θ obtained from ARPRS studies. B and A modes show fourfold and twofold symmetry, respectively. This clearly suggests that $CrPS_4$ exhibits anisotropic phononic behaviour originating from the anisotropic crystal structure, as seen in STEM-HDAAF images Figure 1b, c.

## 2.0 Energy-dispersive X-ray

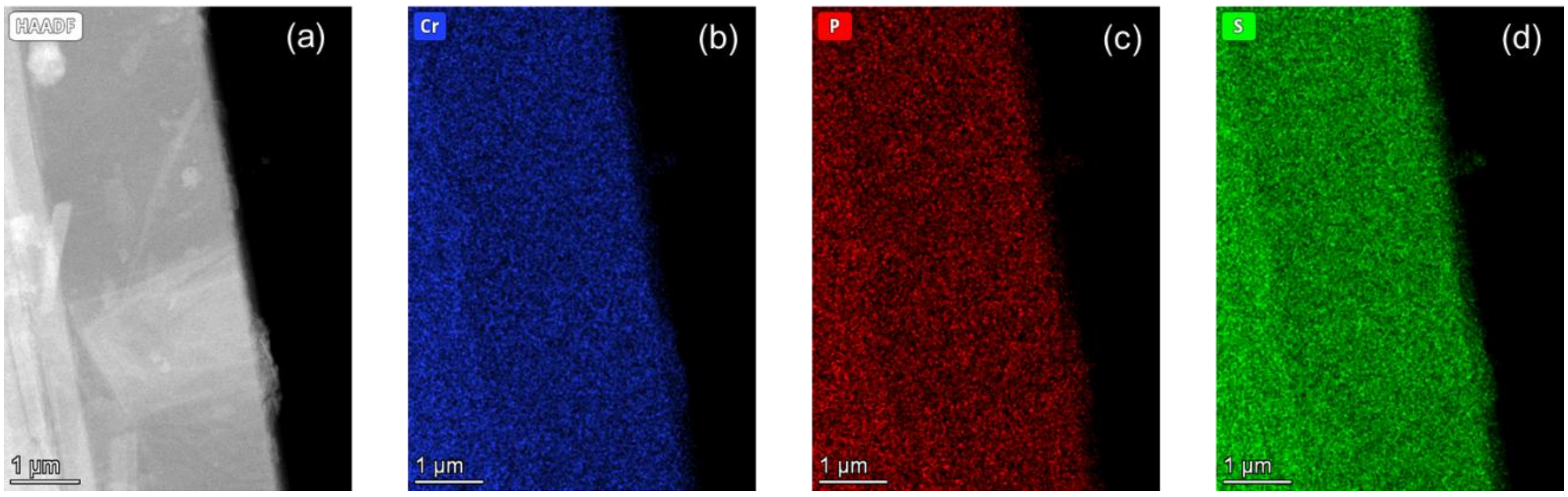

**Figure S2**: EDS mapping (a) optical image of $CrPS_4$ (b) EDS map of elements (Cr, P and S) in (b), (c) and (d), respectively, demonstrating identical contribution of atoms on the entire crystal.

## 3.0 Magnetization study

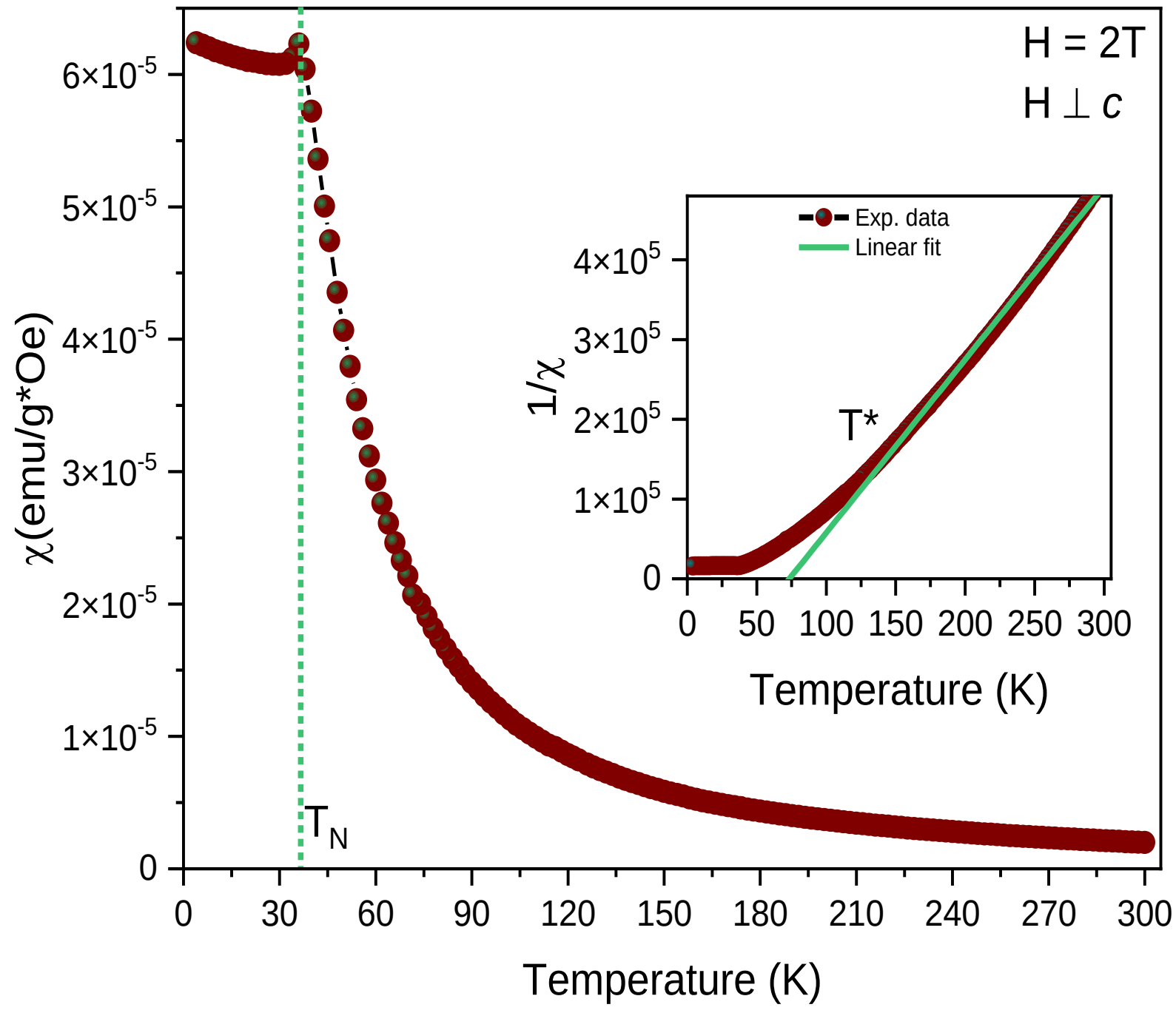

**Figure S3:** Susceptibility χ vs T under magnetic field of 2T applied perpendicular to *c*-axis, green dotted line marks $T_N$ ~ 40K. The inset shows the inverse of susceptibility with T, highlighting deviation at T* from linear curve.

## 4.0 Temperature-dependent Raman spectrum

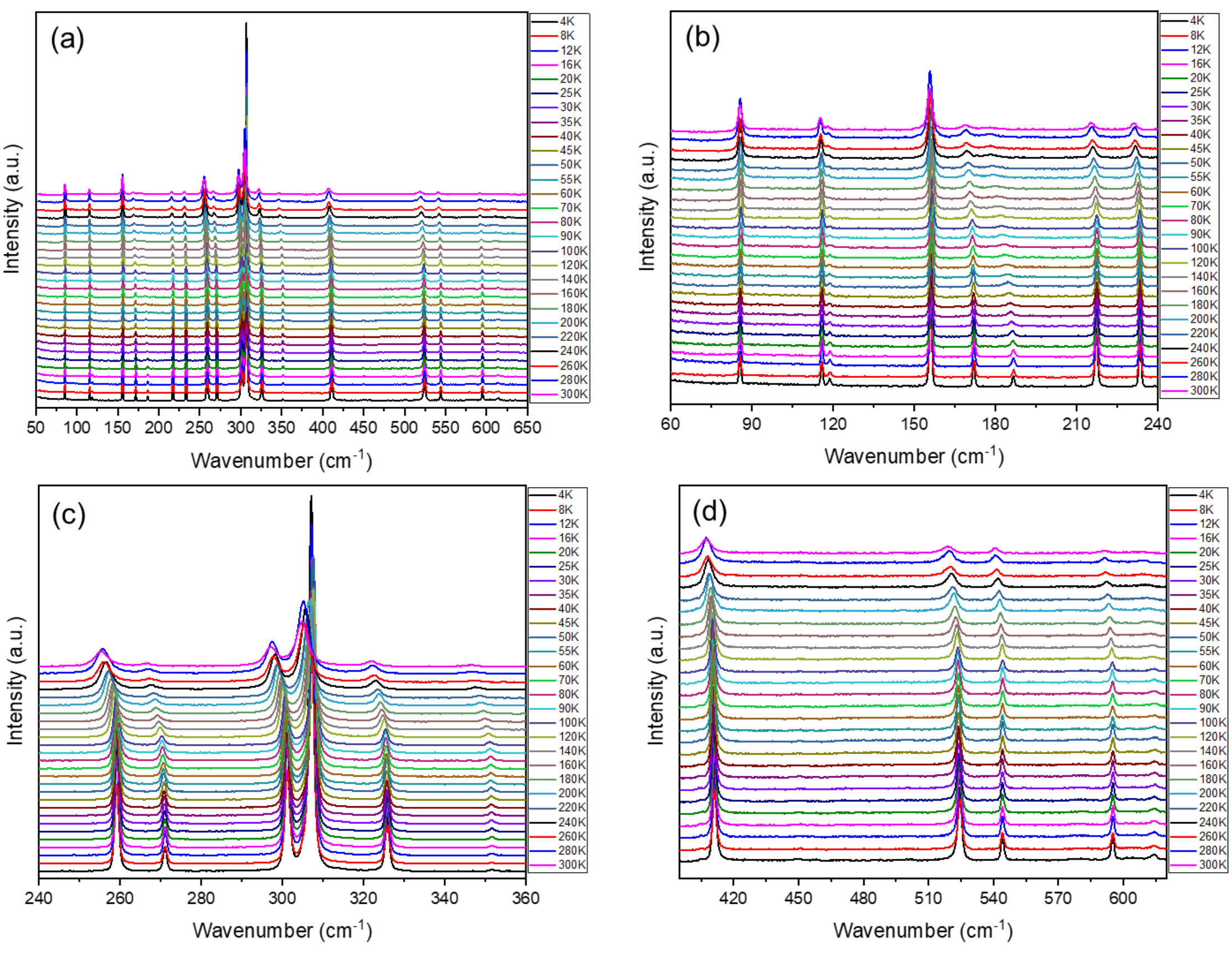

**Figure S4:** (a) Temperature-dependent stack plot of Raman spectra. For clarity, (b), (c), and (d) represent their zoomed versions. The softening of the Raman mode is observed as the temperature rises from 4 K to 300 K

## Table S1: Assigned Raman modes and their corresponding atomic vibration

| Assigned Raman modes | Raman Frequency ($cm^{-1}$) | Atomic vibration of $CrPS_4$ J. Pandey *et al* [28] |
|---|---|---|
| $B_1$ | 84.8 $cm^{-1}$ | Out-of-phase Cr translation, Out-of-phase S displacement |
| $B_2$ | 115 $cm^{-1}$ | In-plane Cr translation, out-of-phase S shearing |
| $A_1$ | 117.7 $cm^{-1}$ | P-S rocking motion |
| $B_3$ | 155 $cm^{-1}$ | Out-of-phase Cr translation, Out-of-phase S displacement |
| $A_2$ | 168.6 $cm^{-1}$ | In-plane Cr-S translation, in-plane P-S translation |
| $A_3$ | 177.4 $cm^{-1}$ | In-plane Cr-S breathing |
| $A_4$ | 215.3 $cm^{-1}$ | In-plane P-S symmetric motion |
| $B_4$ | 231 $cm^{-1}$ | Cr-S octahedra shearing |
| $B_5$ | 255 $cm^{-1}$ | In-plane P-S symmetric motion |
| $A_5$ | 266.5 $cm^{-1}$ | Cr-S octahedra breathing |
| $B_6$ | 297 $cm^{-1}$ | In-plane Cr-S octahedra breathing |
| $A_6$ | 304.5 $cm^{-1}$ | Cr-S octahedra breathing |
| $A_7$ | 321.8 $cm^{-1}$ | In plane Cr-S symmetric motion, in-plane P-S translation |
| $B_7$ | 346.5 $cm^{-1}$ | Out of phase Cr-S stretching, Out-of-phase Cr-S stretching |
| $A_8$ | 407 $cm^{-1}$ | P-S tetrahedra breathing |
| $A_9$ | 519 $cm^{-1}$ | P-S tetrahedra breathing |
| $A_{10}$ | 540.7 $cm^{-1}$ | P-S tetrahedra breathing |
| $B_8$ | 591 $cm^{-1}$ | Out-of-plane P motion with S displacement |
| $A_{11}$ | 607.6 $cm^{-1}$ | ------ |

## 5.0 Temperature-dependent Raman shift

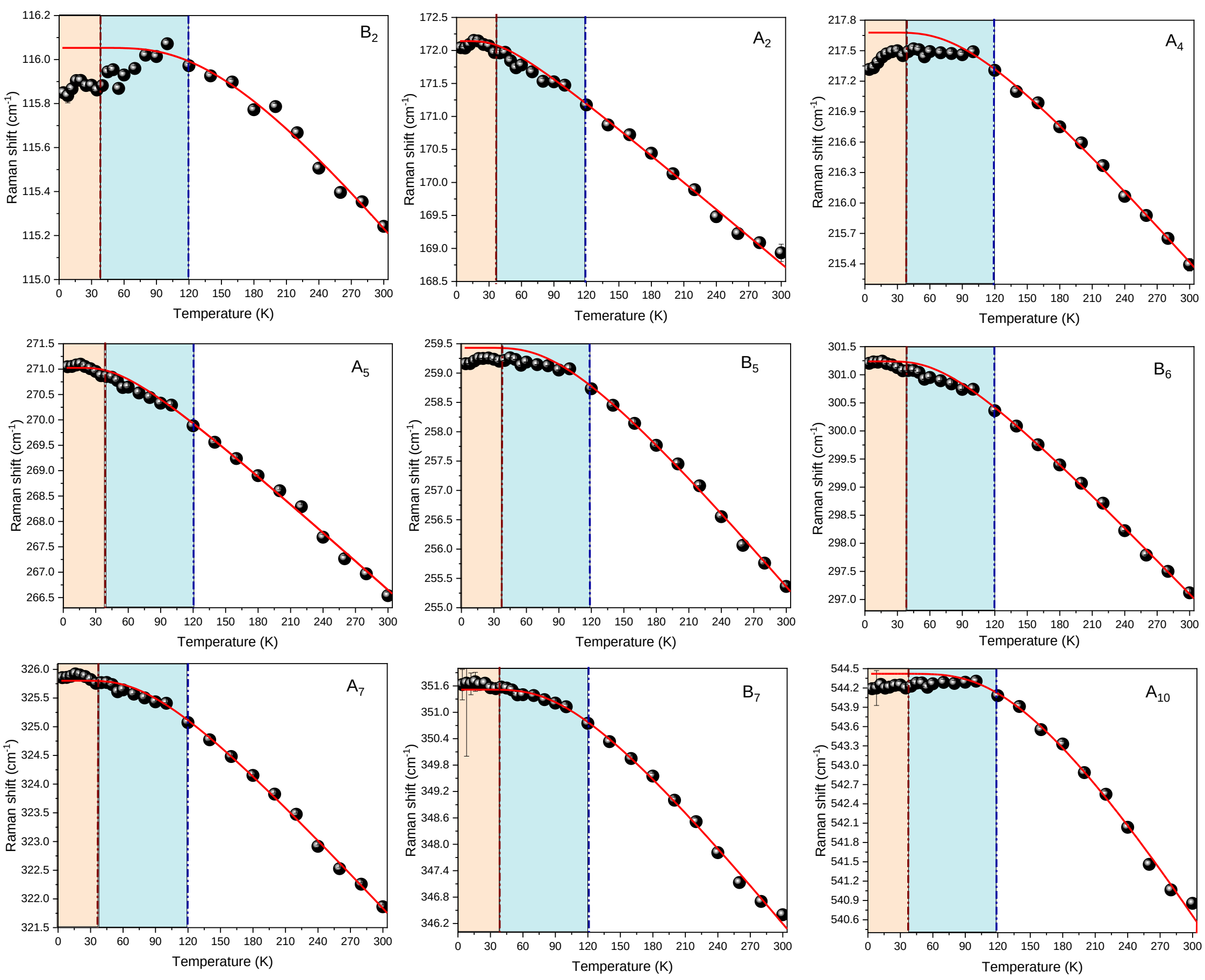

**Figure S5**: T evolution of the Raman shift showing anomalies at T*~120 K as well as the magnetic ordering transition temperature $T_N$.

## 6.0 Spin-phonon coupling

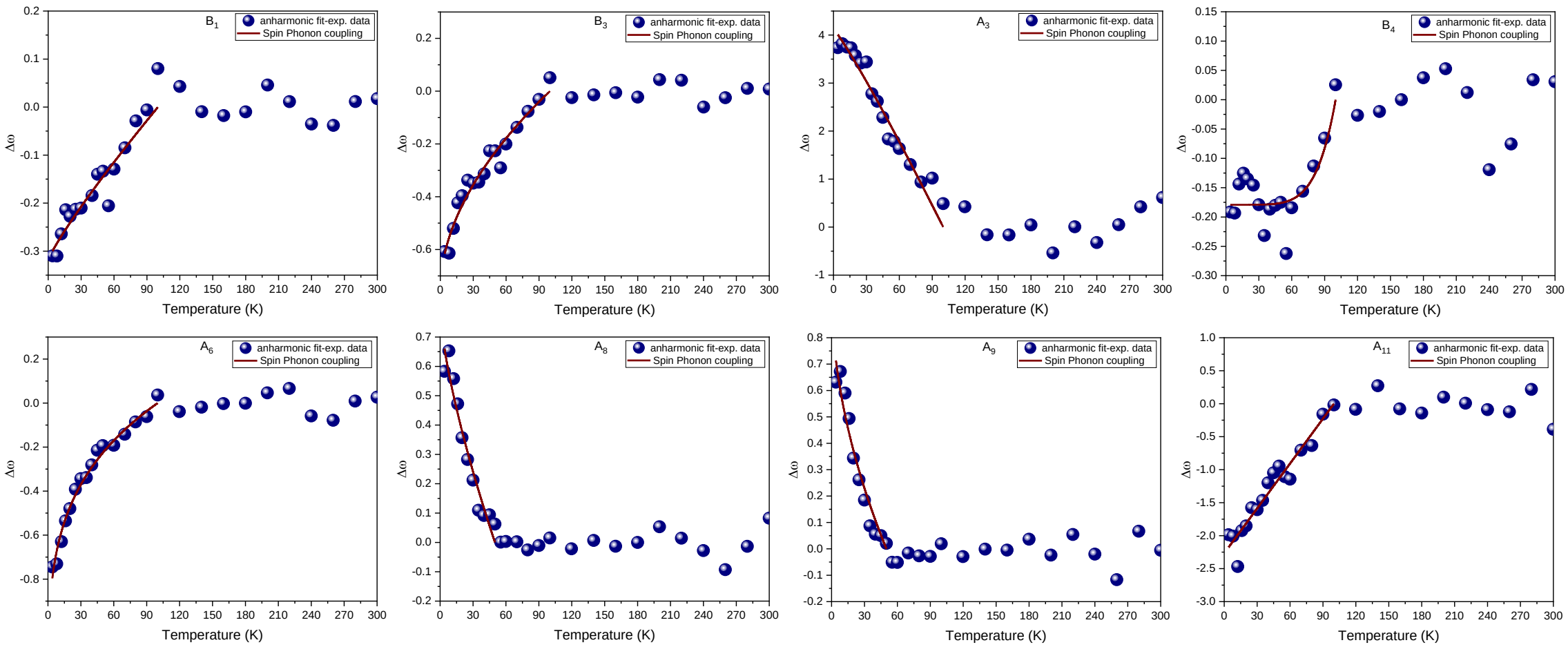

**Figure S6**: These represent *Δω* deviation of experimental data from anharmonic behaviour, and these *Δω* This is due to spin. So, the data were fitted using spin-phonon coupling equations

## 7.0 Temperature-dependent Photoconductivity

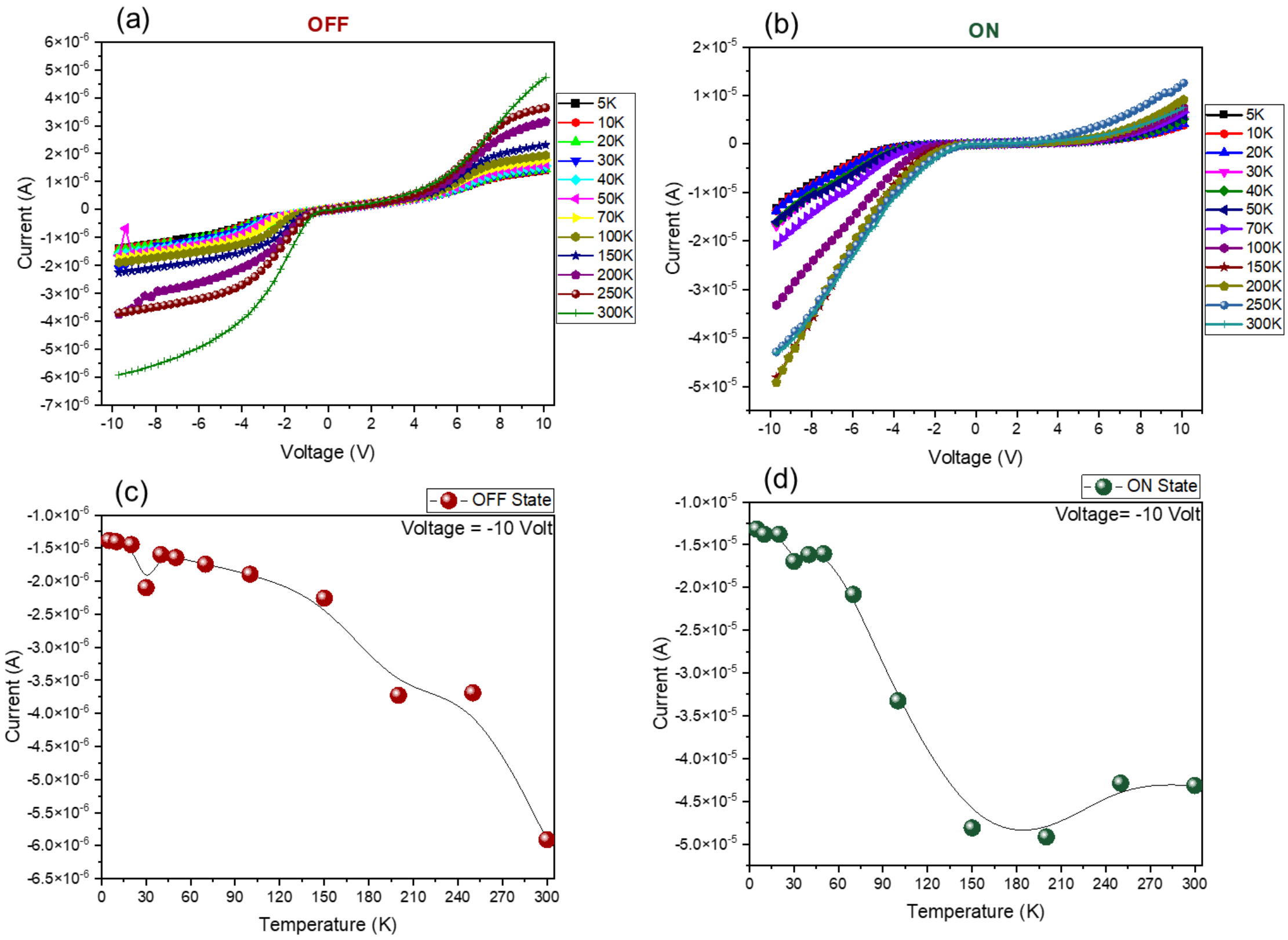

**Figure S7**: (a) & (b) are I-V at different temperatures with light OFF and ON conditions, respectively. (c) &(d) are the current values at a voltage of -10V, and plotted them against temperature.

## 8.0 X-ray absorption spectroscopy

X-ray absorption measurements were done at the Cr L-edge to investigate the changes in the local structural environment of the Cr site in $CrPS_4$. At the Cr L-edge, the absorption occurs from 2p to 3d electronic states. Measurements were done at Room temperature as well as low temperature to identify the temperature-related changes in the local environment of the Cr site, which relate to octahedral distortion as observed in our temperature-dependent Raman measurements. The system is cooled using a cold finger setup, where liquid nitrogen is used as the cooling medium; it was expected that the temperature reached at the sample would be ~ 100 K. Room-temperature XAS were compared with the low-temperature XAS, observing modifications in spectral line shape. This modification at the Cr L-edge with temperature confirms local structural distortion in the $CrS_6$ octahedra in $CrPS_4$.

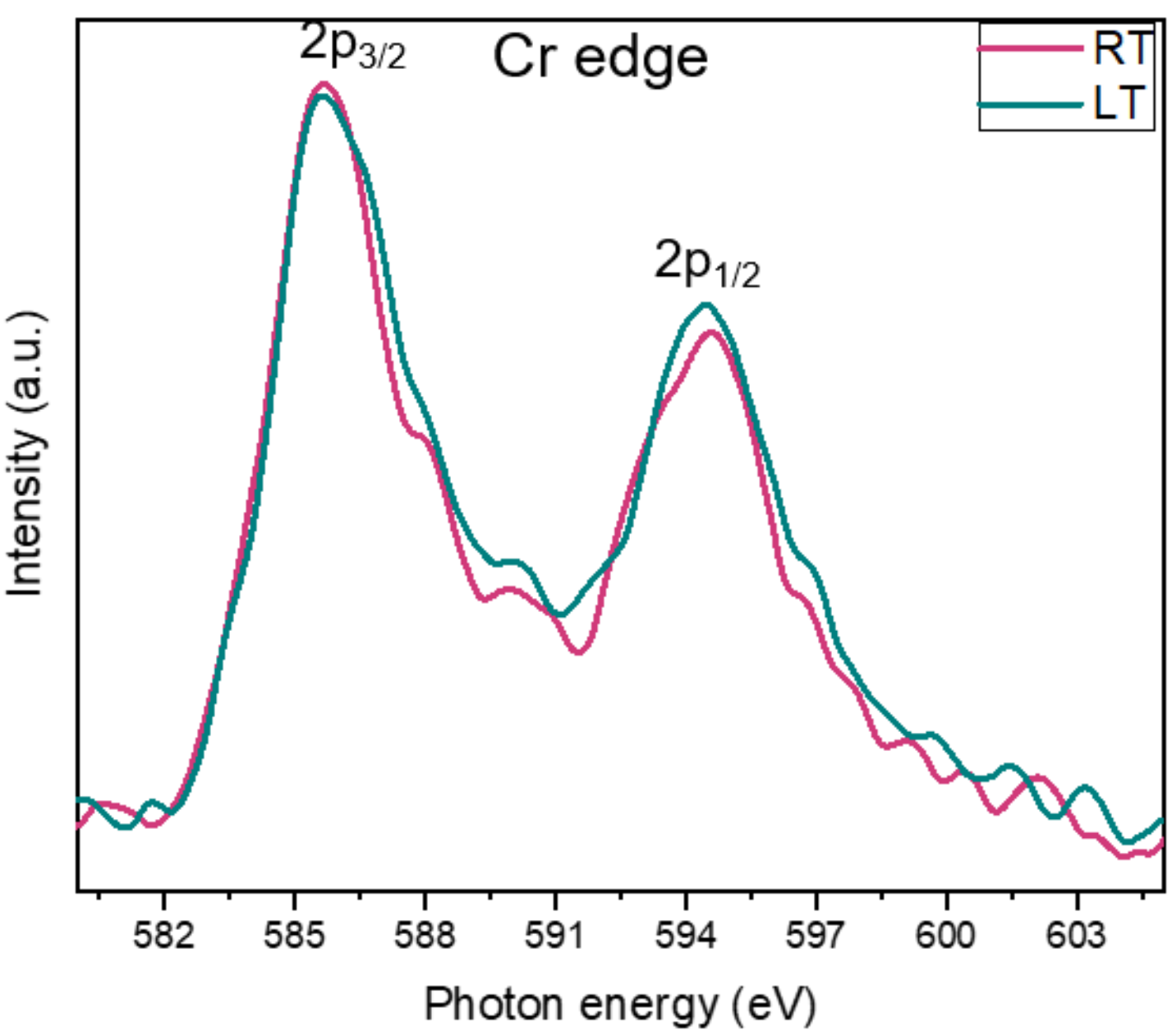

**Figure S8**: Cr L-edge X-ray absorption spectra of $CrPS_4$ at the Cr (L-edge) at RT (room-temperature) and LT (low temperature), revealing temperature-induced modification in $CrS_6$ octahedra.

## 9.0 Raman mapping of Heterostructure

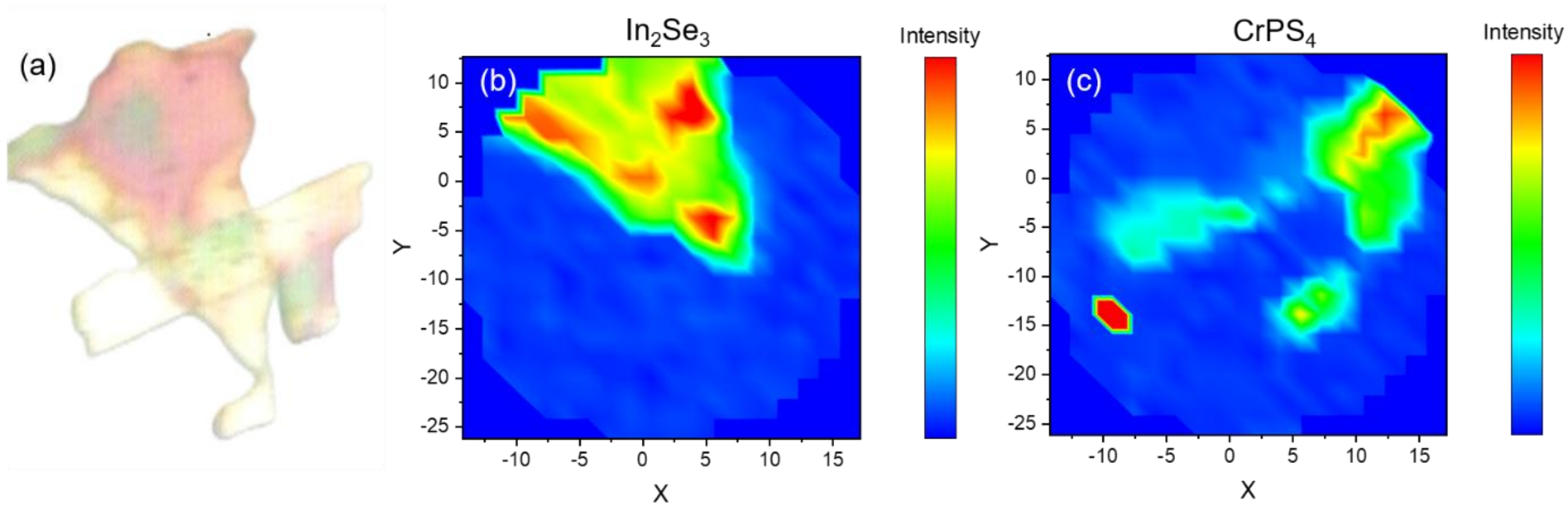

**Figure S9**: Raman mapping of the heterostructure. (a) Optical image (b) Raman mapping showing $In_2Se_3$ area (c) shows $CrPS_4$ Area.

## 10.0 Temperature-dependent Raman spectra of heterostructure

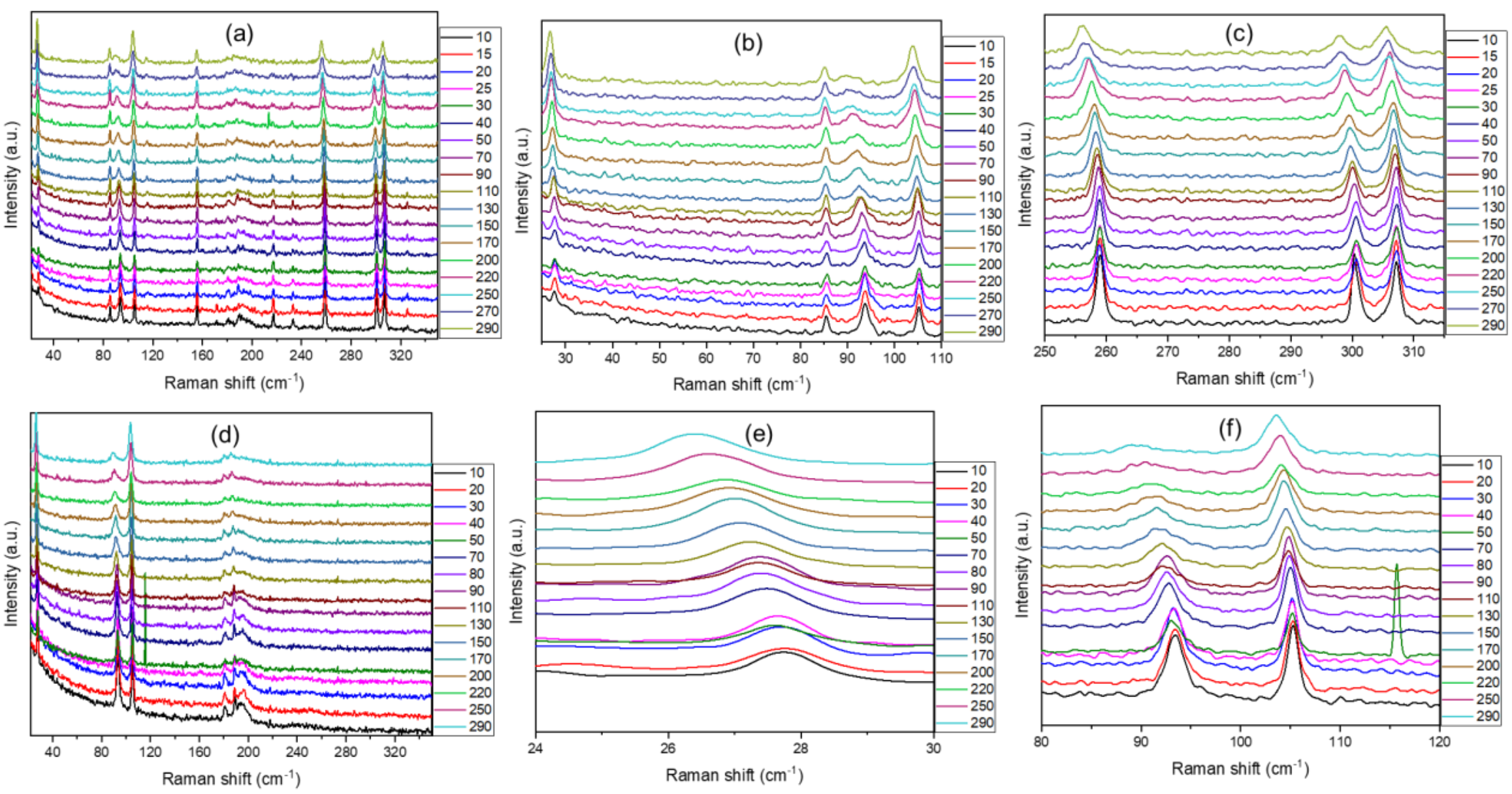

**Figure S10**: Temperature-dependent Raman spectra of $CrPS_4$/ $In_2Se_3$ taken at the point of the red star as mentioned in the main text; (b), (c) are its zoomed versions, (d) stack plot of isolated $In_2Se_3$, (e) and (f) its zoomed version.

## 11.0 Temperature-dependent Photoluminescence

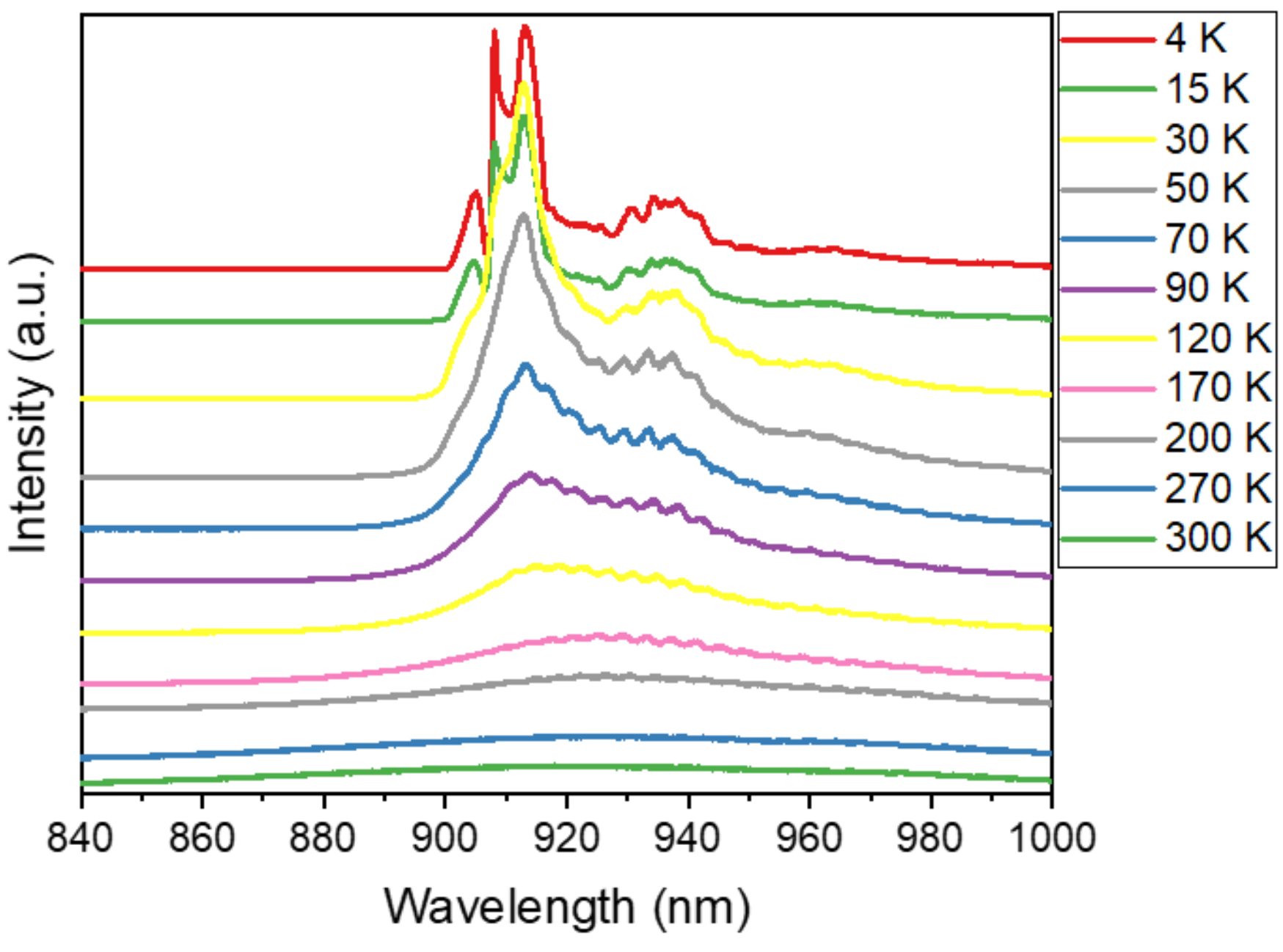

**Figure S11**: Photoluminescence spectrum stack plot showed at different temperatures

## References

[1] S. Jiang, J. Shan, and K. F. Mak, *Electric-field switching of two-dimensional van der Waals magnets*, Nat. Mater. 17, 406–410 (2018).
[2] P. Zhang, T. F. Chung, Q. Li, S. Wang, Q. Wang, W. L. B. Huey, S. Yang, J. E. Goldberger, J. Yao, and X. Zhang, *All-optical switching of magnetization in atomically thin* $CrI_3$*,* Nat. Mater. 21, 1373–1378 (2022).
[3] M. Siskins, S. Kurdi, M. Lee, B. J. M. Slotboom, W. Xing, S. M.Valero, E. Coronado, S. Jia, W. Han, T. Van Der Sar, H. S. J. Van Der Zant, and P. G. Steeneken, *Nanomechanical Probing and Strain Tuning of the Curie Temperature in Suspended Cr2ge2te6-based Heterostructures*, Npj 2D Materials and Applications 6 (2022).
[4] D. Lebedev, J. T. Gish, E. S. Garvey, T. K. Stanev, J. Choi, L. Georgopoulos, T. W. Song, H. Y. Park, K. Watanabe, T. Taniguchi, N. P. Stern, V. K. Sangwan, M. C. Hersam, *Electrical Interrogation of Thickness-Dependent Multiferroic Phase Transitions in the 2D Antiferromagnetic Semiconductor* $NiI_2$. Adv. Funct. Mater. 33, 2212568 (2023).
[5] S. Jiang, L. Li, Z. Wang, K. F. Mak, and J. Shan, *Controlling Magnetism in 2D Cri3 by Electrostatic Doping*, Nature Nanotechnology 13, 549 (2018).
[6] Y. J. Sun, J. M. Lai, S.M. Pang, X. L. Liu, P. H. Tan, J. Zhang*, Magneto-Raman study of magnon–phonon coupling in two-dimensional Ising antiferromagnetic* $FePS_3$,” J. Phys. Chem. Lett. 13, 1533–1539 (2022).
[7] X. Wang, F. Wang, and H. Xiang, *First-principles studies of multiferroic and magnetoelectric materials*, Science Bulletin 60, 156–181 (2015).
[8] N. A. Spaldin and R. Ramesh, *Advances in magnetoelectric multiferroics*, Nature Materials 18, 203–212 (2019).
[9] J. Sinova, S. O. Valenzuela, J. Wunderlich, C. H. Back, and T. Jungwirth, *Spin Hall effects*, Reviews of Modern Physics 87, 1213–1260 (2015).
[10] K. Uchida, S. Takahashi, H. Harii, et al., *Observation of the spin Seebeck effect*, Nature 455, 778–781 (2008).
[11] Y. Liu, N. O. Weiss, X. Duan, H.-C. Cheng, Y. Huang, and X. Duan, *Van der Waals heterostructures and devices*, Nature Reviews Materials 1, 16042 (2016).
[12] Y. Liu, Y. Huang, and X. Duan, *Van der Waals integration before and beyond two-dimensional materials*, Nature 567, 323–333 (2019).
[13] R. Lan, X. Luo, N. Zhou, A. Wang, M. Cheng, L. Liu, Y. Pan, R. Zhang, J, Li, *Thermomagnetic irreversibility in a* $Cr_{1.45}Te_2$ *crystal: Role of spin-phonon coupling***,** Phys. Rev. B 112, 104414 (2025).
[14] E. Ergeçen, B. Ilyas, J. Kim, J. Park, M. B. Yilmaza, T. Luo, D. Xiao, S. Okamoto, J. G. Park, and N. Gedik, *Coherent detection of hidden spin–lattice coupling in a vander Waals antiferromagnet*, PNAS 120 (2023).
[15] C. H. Sohn, C. H. Kim, L. J. Sandilands, N. T. M. Hien, S. Y. Kim, H. J. Park, K. W. Kim, S. J. Moon, J. Yamaura, Z. Hiroi, and T. W. Noh, *Strong Spin-Phonon Coupling Mediated by Single Ion Anisotropy in the All-In-All-Out Pyrochlore Magnet* $Cd_2Os_2O_7$*,* Phys. Rev. Lett. 118, 117201 (2017).
[16] B. Sadhukhan, A. Bergman, Y. O. Kvashnin, J. Hellsvik, and A. Delin, *Spin-lattice Couplings in Two-dimensional Cri3 from First-principles Computations*, Physical Review B 105 (2022).
[17] D. P. Kozlenko, O. N. Lis, S. E. Kichanov, E. V. Lukin, N. M. Belozerova, and B. N. Savenko, *Spin-induced Negative Thermal Expansion and Spin–phonon Coupling in Van Der Waals Material CrBr3*, Npj Quantum Materials 6, (2021).
[18] L. Zhu, X. Zeng, H. Luo, H. Shang, and Z. Li, *Theoretical Raman Study of In-plane Spin–phonon Coupling in a Crcl3 Monolayer*, The Journal of Physical Chemistry C 129, 9954 (2025).
[19] X. Guo, Q. Feng, Z. Tian, A. Shen, M. M. Al-Makeen, Y. Wang, H. Xie, Y. Nie, Q. Xia, and H. Huang, *Spin–phonon Coupling in Two-dimensional Antiferromagnet Fe0.25tase2*, Applied Physics Letters 128, (2026).
[20] X. Chen, X. Zhang, W. He, Y. Li, J. Lu, D. Yang, D. Li, L. Lei, Y. Peng, and G. Xiang, *Lattice Dynamics and Phonon Dispersion of the Van Der Waals Layered Ferromagnet Fe3Gate2*, Nano Letters 25, 4353 (2025).
[21] Q.-Q. Ye, K. Liu, and Z.-Y. Lu, *Influence of Spin-phonon Coupling on Antiferromagnetic Spin Fluctuations in FeSe Under Pressure: First-principles Calculations with Van Der Waals Corrections*, Physical Review B 88, (2013).
[22] L. Hu, K. Z. Du, Y. Chen, *Spin-phonon coupling in two-dimensional magnetic materials*. National Science Open, 2(4): 20230002 (2023).
[23] J. Son, S. Son, P. Park, M. Kim, Z. Tao, J. Oh, T. Lee, S. Lee, J. Kim, K. Zhang, K. Cho, T. Kamiyama, J. Hee Lee, K. F. Mak, J. Shan, M. Kim, J. G. Park, J. Lee, *Air-Stable and Layer-Dependent Ferromagnetism in Atomically Thin van der Waals* $CrPS_4$, ACS Nano 15 (10) (2021).
[24] Y. Peng, S. Ding, M. Cheng, Q. Hu, J. Yang, F. Wang, M. Xue, Z. Liu, Z. Lin, M. Avdeev, Y. Hou, W. Yang, Y. Zheng, and J. Yang, *Magnetic Structure and Metamagnetic Transitions in the Van Der Waals Antiferromagnet* $CrPS_4$, Advanced Materials 32, 2001200 (2020).
[25] S. Calder, A. V. Haglund, Y. Liu, D. M. Pajerowski, H. B. Cao, T. J. Williams, V. O. Garlea, and D. Mandrus, *Magnetic Structure and Exchange Interactions in the Layered Semiconductor* $CrPS_4$, Physical Review B 102, (2020).
[26] V. Multian, F. Wu, D. Van Der Marel, N. Ubrig, and J. Teyssier, *Brightened Optical Transition Hinting to Strong Spin-lattice Coupling in a Layered Antiferromagnet*, Advanced Science 12, (2025).

[27] P. Gu, Q. Tan, Y. Wan, Z. Li, Y. Peng, J. Lai, J. Ma, X. Yao, S. Yang, K. Yuan, D. Sun, B. Peng, J. Zhang, and Y. Ye, *Photoluminescent Quantum Interference in a Van Der Waals Magnet Preserved by Symmetry Breaking*, ACS Nano 14, 1003 (2020).
[28] J. Pandey, D. Wulferding, W. Cho, M.-C. Jung, J. Jeong, N. Myoung, M. J. Han, S. Cho, and H. Yang, *Spin–phonon Coupling and Magnetic Ordering in Layered $CrPS_4$*, ACS Applied Materials & Interfaces 18, 19797 (2026).
[29] T. Fąs, M. Wlazło, M. Birowska, M. Rybak, M. Zinkiewicz, L. Oleschko, M. Goryca, Ł. Gondek, B. Camargo, J. Szczytko, A. K. Budniak, Y. Amouyal, E. Lifshitz, and J. Suffczyński, *Direct Optical Probing of the Magnetic Properties of the Layered Antiferromagnet $CrPS_4$*, Advanced Optical Materials 13, (2025).
[30] A. K. Budniak, N. A. Killilea, S. J. Zelewski, M. Sytnyk, Y. Kauffmann, Y. Amouyal, R. Kudrawiec, W. Heiss, and E. Lifshitz, *Exfoliated $CrPS_4$ with Promising Photoconductivity*, Small 16, 1905924 (2020).
[31] S. Kim, S. Yoon, H. Ahn, G. Jin, H. Kim, M.-H. Jo, C. Lee, J. Kim, and S. Ryu, *Photoluminescence Path Bifurcations by Spin Flip in Two-dimensional $CrPS_4$*, ACS Nano 16, 16385 (2022).
[32] J. Yun, S. Son, J. Shin, G. Park, K. Zhang, Y. J. Shin, J.-G. Park, and D. Kim, *Magnetic Proximity-induced Superconducting Diode Effect and Infinite Magnetoresistance in a Van Der Waals Heterostructure*, Physical Review Research 5, (2023).
[33] D. K. de Wal, A. Iwens, T. Liu, P. Tang, G. E. W. Bauer, and B. J. van Wees, *Long-distance magnon transport in the van der Waals antiferromagnet $CrPS_4$*, Phys. Rev. B 107, (2023).
[34] J. Chen, X. Xie, S. Li, Z. Liu, J.-T. Wang, J. He, and Y. Liu, *Layer-resolved Ferromagnetic and Antiferromagnetic Proximity Effects in Crps4/wse2 Heterostructures*, The Journal of Physical Chemistry Letters 16, 10720 (2025).
[35] S. Maity, D. Dey, A. Ghosh, S. Masanta, B. K. De, H. S. Kunwar, B. Das, T. Kundu, M. Palit, S. Bera, K. Dolui, K. Watanabe, T. Taniguchi, L. Yu, A. Taraphder, and S. Datta, *Manipulating Spin-lattice Coupling in Layered Magnetic Topological Insulator Heterostructure via Interface Engineering*, Advanced Functional Materials 34, (2024).
[36] W. Ding, J. Zhu, Z. Wang, Y. Gao, D. Xiao, Y. Gu, Z. Zhang, and W. Zhu, *Prediction of Intrinsic Two-dimensional Ferroelectrics in In2Se3 and Other III2-VI3 Van Der Waals Materials*, Nature Communications 8, 14956 (2017).
[37] H. Li, W. Zhu, *Spin-Driven Ferroelectricity in Two-Dimensional Magnetic Heterostructures,* Nano Lett. 23 (22), 10651–10656 (2023).
[38] D. M. Phase, M. Gupta, S. Potdar, L. Behera, R. Sah and A. Gupta, *Development of soft X-ray polarized light beamline on Indus-2 synchrotron radiation source,* AIP Conf. Proc. 1591, 685–686 (2014).
[39] K. Momma, F. Izumi, *Vesta: a three-dimensional visualisation system for electronic and structural analysis*, Applied Crystallography 41 (3) 653–658 (2008).
[40] M. H. Nguyen, S. Son, G. Park, W. Na, J.-G. Park, and H. Cheong, *Temperature-Dependent Raman Study of Antiferromagnetic $CrPS_4$*, J. Mater. Chem. C 12, 12468 (2024).
[41] S. Kim, J. Lee, C. Lee, and S. Ryu, *Polarized Raman Spectra and Complex Raman Tensors of Antiferromagnetic Semiconductor $CrPS_4$*, The Journal of Physical Chemistry C 125, 2691 (2021).
[42] J. Lee, T. Y. Ko, J. H. Kim, H. Bark, B. Kang, S.-G. Jung, T. Park, Z. Lee, S. Ryu, and C. Lee, *Structural and Optical Properties of Single- and Few-layer Magnetic Semiconductor $CrPS_4$*, ACS Nano 11, 10935 (2017).
[43] M. Li, X. Wei, Q. Xie, L. Chen, L. Ma, and G. Cheng, *Investigation on the Intrinsic Phonon Properties of Crps4: A Combined Raman Spectroscopy and First-principles Calculations Study*, ACS Omega 10, 31179 (2025).
[44] Pankaj K. Pandey, R. J. Choudhary, Dileep K. Mishra, V. G. Sathe, and D. M. Phase. APPLIED PHYSICS LETTERS 102, 142401 (2013).
[45] M. Balkanski, R. F. Wallis, and E. Haro, *Anharmonic effects in light scattering due to optical phonons in silicon*, Physical Review B 28, (1983).
[46] P.G. Klemens, *Anharmonic decay of optical phonons*, Phys. Rev., 148 (1966).
[47] A. Ghosh, M. Palit, S. Maity, V. Dwij, S. Rana, and S. Datta, *Spin-phonon coupling and magnon scattering in few-layer antiferromagnetic FePS3,* Physical Review B 103(6), (2021).
[48] E. Granado, A. Garcia, J. A. Sanjurjo, C. Rettori, and I. Torriani, *Magnetic ordering effects in the Raman spectra of $La_{1-x}Mn_{1-x}O_3$*, PHYSICAL REVIEW B 60 (17), (1999).
[49] D. J. Lockwood, M. G. Cottam, *The spin-phonon interaction in $FeF_2$ and $MnF_2$ studied by Raman spectroscopy*, J. Appl. Phys. 64, 5876–5878 (1988).
[50] Hung CH, Shih PH, Wu FY, Li WH, Wu SY, Chan TS, Sheu HS. *Spin-phonon coupling effects in antiferromagnetic Cr2O3 nanoparticles*. Journal of Nanoscience and Nanotechnology 10(7) (2010).
[51] B. S. Araújo, A. M. Arévalo-López, C. C. Santos, J. P. Attfield, C. W. A. Paschoal, A. P. Ayala, *Spin–phonon coupling in monoclinic $BiCrO_3$*, Journal of Applied Physics 127 (2020).
[52] Q. L. Pei, X. Luo, G. T. Lin, J. Y. Song, L. Hu, Y. M. Zou, L. Yu, W. Tong, W. H. Song, W. J. Lu, and Y. P. Sun, *Spin Dynamics, Electronic, and Thermal Transport Properties of Two-dimensional Crps4 Single Crystal*, Journal of Applied Physics 119, 043902 (2016).

[53] R. A. Susilo, B. G. Jang, J. Feng, Q. Du, Z. Yan, H. Dong, M. Yuan, C. Petrovic, J. H. Shim, D. Y. Kim, and B. Chen, *Band gap crossover and insulator–metal transition in the compressed layered $CrPS_4$,* npj Quantum Mater. 5, 58 (2020).
[54] Y. Ohno, A. Mineo, I. Matsubara, *Reflection electron-energy-loss spectroscopy, x-ray-absorption spectroscopy, and x-ray photoelectron spectroscopy studies of a new type of layer compound CrPS4,* Phys. Rev. B **40**, 10262 (1989).
[55] M. Riesner, A. K. Budniak, Y. Amouyal, E. Lifshitz, and G. Bacher, *Temperature dependence of Fano resonances in $CrPS_4$,* The Journal of chemical physics, 156 (2022).
[56] K. P. O. Donnell, X. Chen, *Temperature dependence of semiconductor band gaps*, Appl. Phys. Lett. 58, 2924 (1991).
[57] Y. Zhang, *Applications of Huang–Rhys theory in semiconductor optical spectroscopy*, Journal of Semiconductors 40, (2019).
[58] D. Jangra, B. K. De, P. Sharma, K. Chakraborty, S. Parate, A. K. Yogi, R. Mittal, M. K. Gupta, P. Nukala, P. K. Velpula, and V. G. Sathe, *Anisotropic Light–matter Interaction in A-in2se3: Wavelength-dependent Study*, ACS Applied Materials & Interfaces 17, 22903 (2025).
[59] R. Baral *et al*, *Correlation-driven spin reorientation via competing anisotropy channels in $CrPS_4$*, arXiv:2605.16541 (2026), https://doi.org/10.48550/arXiv.2605.16541
[60] G. Buccoliero *et al, Ligand-mediated magnetoelectronic coupling across metamagnetic transitions in $CrPS_4$,* arXiv:2602.16083 (2026)
[61] Stavitski, Eli. et al. *The CTM4XAS program for EELS and XAS spectral shape analysis of transition metal L edges*. Micron 41 (2010) 687-694.
[62] K. Huang and A. Rhys, *Theory of Light Absorption and Non-radiative Transitions in F -centres*, Proceedings of the Royal Society of London. A. Mathematical and Physical Sciences 204, 406 (1950).